\documentclass[pra,amssymb,amsmath,twocolumn,floatfix]{revtex4}

\usepackage[english]{babel}

\usepackage{amsthm}
\usepackage{amsmath}
\usepackage{amssymb}
\usepackage{color}
\usepackage[dvips]{graphicx}
\usepackage{psfrag}

\newcommand{\ra}{\rangle}
\newcommand{\la}{\langle}
\newcommand{\be}{\begin{equation}}
\newcommand{\ee}{\end{equation}}
\newcommand{\bea}{\begin{eqnarray}}
\newcommand{\eea}{\end{eqnarray}}
\newcommand{\ba}{\begin{array}}
\newcommand{\ea}{\end{array}}
\newcommand{\ben}{\begin{enumerate}}
\newcommand{\een}{\end{enumerate}}
\newcommand{\bi}{\begin{itemize}}
\newcommand{\ei}{\end{itemize}}

\newcommand{\bt}{\begin{table}}
\newcommand{\et}{\end{table}}
\newcommand{\btr}{\begin{tabular}}
\newcommand{\etr}{\end{tabular}}
\newcommand{\bfi}{\begin{figure}}
\newcommand{\efi}{\end{figure}}

\newcommand{\nn}{\nonumber}

\begin{document}

\title{Mapping all classical spin models to a lattice gauge theory}

\author{G.~De las Cuevas, W.~D\"ur, and H.~J.~Briegel}
\affiliation{
Institut f{\"u}r Theoretische Physik, Universit{\"a}t
Innsbruck, Technikerstra{\ss}e 25, A-6020 Innsbruck,
Austria\\
Institut f\"ur Quantenoptik und Quanteninformation der \"Osterreichischen Akademie der Wissenschaften, A-6020 Innsbruck, Austria}

\author{M.~A.~Martin--Delgado}
\affiliation{
Departamento de F\'{\i}sica Te\'orica I, Universidad Complutense, 28040 Madrid, Spain}



\maketitle

\tableofcontents

\section{Introduction}
\label{sec:introduction}

Usually, classical spin models have \emph{global} symmetries, that is, they are invariant under a transformation which is the same at all points in spacetime.
Instances of such theories are Newtonian dynamics under Galilean transformations, special relativity under Lorentz transformations, or the choice of the zero point of energies in any system. These symmetries are motivated by the homogeneity and isotropy of space, and the homogeneity of time, which render the choice of the origin of the coordinate system, as well as the orientation of its axis, arbitrary.
The same sort of symmetries apply for usual statistical models, such as the Ising or Potts model, where a global rotation of the classical spins leaves the physics of the system invariant~\cite{Wu84}. We will henceforth refer to these models as Standard Statistical Models (SSMs). 

The notion of global symmetry can be lifted to the notion of \emph{local} symmetry, where the applied transformation is point--dependent.
This is precisely the case for gauge theories, which describe the most fundamental interactions in nature, like quantum electrodynamics (QED), weak interactions, and quantum chromodynamics (QCD). These symmetries are physically motivated by the conservation of certain quantities at every point in spacetime. More precisely, the conservation of electric charge, the weak neutral current, and the color charge induce the $U(1)$, $SU(2)$, and $SU(3)$ gauge symmetries in QED, weak interaction theory, and QCD, respectively. 

Lattice gauge theories (LGTs) are lattice formulations of gauge theories~\cite{Ko79}. That is, they are theories in which the variables sit at the edges of a lattice, and the Hamilton function of the system exhibits some local symmetries. As we will elaborate below, they are interesting not only as a means to study gauge theories, but also as a new class of statistical models \emph{per se}.

Given this variety of models (SSMs, with global symmetries, and LGTs with local symmetries), one may wonder if it is possible to relate them, that is, to find a mapping  such that having information of one model automatically yields insight into another model. The purpose of this paper is to show that a general mapping of this kind exists. More precisely, we will prove that the partition function of any (Abelian, discrete) classical spin model can be expressed as the partition function of one specific model, which is \emph{enlarged} and which contains \emph{inhomogeneous} coupling strengths. 
Note that the partition function is the crucial quantity of a model, since, for SSMs, one can obtain all thermodynamical properties by taking derivatives thereof~\cite{Pa}, and for LGTs, one can compute relevant quantities such as Wilson loops~\cite{It89,Cr83,Mo84}.
Following Ref.~\cite{Va08}, we will call this a \emph{completeness} result, since it shows that the partition function of one specific model can specialize (by being defined on a large enough lattice, and by tuning in it the appropriate coupling strengths) to the partition function of any other (Abelian, discrete) classical spin model.

Similar completeness results have been proven before. A very general result was proven in Ref.~\cite{Va08}, where it was shown that the 2D Ising model with magnetic fields is complete for all other (Abelian, discrete) classical spin models. However, the complete model needed to allow for \emph{complex} coupling strengths in order to specialize to other models, and thereby hindered a physical interpretation of the complete model. In Ref.~\cite{De08} several completeness results with real parameters were obtained, and it was shown, e.g.~that the 3D Ising model is complete with real parameters for all other Ising models with fields. 
While similar results for other types of models were shown, none was as general as the original result of~\cite{Va08}, thereby giving the impression that there was a tradeoff between the real parameters and the generality of the completeness results. 
This tradeoff has proven to be false, for we showed in Ref.~\cite{De09} that one can obtain a complete model for all others (Abelian, discrete) classical spin models employing only real parameters.
Moreover, we showed that the Hamilton function of a subsystem of the complete model equals the Hamilton function of any given classical spin model.

One may wonder why LGTs are interesting from the point of view of the completeness results.
The reason is that gauge theories are models with local symmetries, which are a generalization of systems with global symmetries. Thus, there are chances to obtain more powerful completeness results than with SSMs, and this is precisely what we have found.

In this work, we elaborate on Ref.~\cite{De09} and illustrate that result with specific examples and constructions. More precisely, in \cite{De09} it was proven that \emph{the partition function of any Abelian, discrete classical spin model can be a mapped onto the partition function of a lattice gauge theory on a 4--dimensional square lattice with gauge group $\mathbb{Z}_2$ (the 4D $\mathbb{Z}_2$ LGT)}. Here, we go beyond Ref.~\cite{De09} in the following points:
\begin{itemize}
\item		
We present a quantum formulation of the partition function, where the latter is expressed as the scalar product between two states (Sec.~\ref{ssec:quantum-formulation}).
Here, we show that one of these two states is a stabilizer state, which reveals some symmetries of the partition function;

\item		
We present a detailed analysis of the merge and deletion rule (Sec.~\ref{ssec:merge-deletion-rules}), the basic tools used to manipulate and transform one interaction pattern to another;

\item
We present a rigorous proof of the method to obtain general $n-$body interactions (Sec.~\ref{ssec:method}); 

\item	
We provide the explicit construction of interaction patterns where all possible $k$--body Ising--type interactions are present (termed $k$--cliques), for all  $k=1,\ldots,n$; in particular, we show how particles propagate and their paths turn around in the lattice, how one can construct 4-- and 5--body Ising--type interactions, and the explicit layout of all $k$--cliques in the 4D lattice (Sec.~\ref{ssec:explicit-construction});

\item	
In Sec.~\ref{ssec:main-result} we give a detailed explanation of the class of models embraced by the main result of this work;

\item
We provide a new discussion of the main result in terms of the space of all theories (Sec.~\ref{ssec:space-of-all-theories}).
\end{itemize}

Moreover, we will explore some of the consequences of this result, going beyond \cite{De09} in the following aspects:
\begin{itemize}
\item
In Sec.~\ref{sec:computing-observables-complete} we give a general observation concerning all the completeness results;

\item
We illustrate our result by explicitly computing the Wilson loop of a lattice gauge theory, and the magnetization of the 2D Ising model from the partition function of the 4D $\mathbb{Z}_2$ LGT (Sec.~\ref{sec:applications}). 
We also discuss how one can map models with global symmetries to a model with local symmetries. We also give an optimized construction of the 2D Ising model.

\item
In Sec.~\ref{ssec:mean-field-theory} we provide an explicit construction of the 4--clique required to compute the mean--field theory.
\end{itemize}

This paper is structured as follows. First, we will present a brief introduction on lattice gauge theories for the non--expert reader (Sec.~\ref{sec:background-lattice-gauge-theories}). In Sec.~\ref{sec:completeness-4D-Z2-LGT} we will prove our main result, namely that the partition function of any Abelian discrete SSM and any Abelian discrete LGT can be mapped to the partition function of the 4D $\mathbb{Z}_2$ LGT.
In Sec.~\ref{sec:computing-observables-complete} we will make a general observation on how to compute observables of a model from the partition function of the complete model. 
 Then we will explore some applications of this result in Sec.~\ref{sec:applications}. In Sec.~\ref{sec:implications} we will focus on two implications of the main result, namely the computational complexity of the 4D $\mathbb{Z}_2$ LGT and a new mean--field theory for $\mathbb{Z}_2$ LGTs. Finally, in Sec.~\ref{sec:extensions} we will present some extensions of the main result, that is, we will show the completeness of other models. In Sec.~\ref{sec:conclusions} we will summarize our results, discuss a possible reduction of our result to 3D and give an outlook on the subject.

\section{Background on lattice gauge theories}
\label{sec:background-lattice-gauge-theories}

LGTs were originally introduced by Wegner with the goal of obtaining statistical models with a vanishing magnetization in all phases, but nonetheless with a non trivial phase diagram~\cite{We71}. In this approach, LGTs are seen as statistical models with local symmetries, which exhibit novel features compared to SSMs. For example, because local symmetries cannot be broken by local order parameters (such as the magnetization)~\cite{El75}, one needs to define non--local order parameters to witness the different phases of these models. 
An example of a global order parameter is the Wilson loop, which is a product of spins sitting at the edges (or, more generally, of gauge fields) over a closed path: $W(C)=\prod_{i\in C}s_i$. 

The Wilson loop was introduced to distinguish confined from deconfined (or Higgs) phases, since in the former phase it exhibits an area law, whereas in the latter it exhibits a perimeter law~\cite{Wi74}. It is the order parameter for LGTs, and it plays a similar role as the magnetization for SSMs.  Another important quantity is the 't Hooft loop, which is related to the Wilson loop via a certain duality transformation.
Physically, the Wilson loop is an order variable, while the 't Hooft loop is a disorder variable, generalizing concepts similar to those in SSMs, which arise in the context of the Kramers-Wannier duality.
In the Hamiltonian formulation of an LGT \cite{Ko75}, Wilson loops on a closed curve $C$, $W(C)$, and 't Hooft loops on a closed curve $C'$, $\tilde{W}(C')$, are operators that satisfy a loop algebra which is a type of Weyl commutation relation, like for momentum and position operators:
\be
W(C) \tilde{W}(C') = z^{L(C,C')}  \tilde{W}(C') W(C)
\ee
where $L(C,C')$ is the linking number of the two loops and $z$ is an element (phase) of the group $\mathbb{Z}_q$, which is the center of the gauge group. 
With this algebra it is possible to draw conclusions about the perimeter and area laws for both types of loops, and use them to characterize phases in LGTs.

The notion of gapped versus gapless phases is also present in LGTs, but gapped phases have a richer structure than in SSMs. In fact, both confined and Higgs phases are gapped, and this is why the Wilson loop is needed to distinguish them.

A correlation length $\xi$, and thus a gap, can be defined by means of a two--point correlation function $G_c(f_1,f_2)$ between elementary faces $f_1$ and $f_2$
separated a distance $l$ apart, as follows:
\be
G_c(l):=\la s_{f_1} s_{f_2}\ra - \la s_{f_1} \ra \la s_{f_2} \ra 
\ee
where the subscript stands for the connected component of the Green function. Thus, when the system is gapped, we find a behaviour $G_c(l) \sim e^{-l/\xi}$, typical of a system with a finite correlation length. The system is gapless when the correlation length becomes infinite.

Now, with the aid of the correlation length and the Wilson and 't Hooft loops, we can give a more precise picture of the more relevant phases in an LGT~\cite{tH78,Ko99}
\begin{description}
\item Confined phase: it is gapped and the Wilson loop obeys an area law, while the 't Hooft loop follows a perimeter law.

\item Deconfined (Higgs) phase: 
it is gapped and the Wilson loop obeys a perimeter law, while the 't Hooft loop follows an area law.

\item Coulomb phase: it is gapless and both loops obey a perimeter law.
\end{description}

In another approach to LGTs, Wilson independently introduced a more general class of LGTs, namely non--Abelian LGTs (as will be defined below), as gauge theories formulated on a discrete spacetime~\cite{Wi74}. In this context, LGTs are cutoff regulations of gauge theories of strongly interacting particles. 
Gauge theories themselves describe three of the four fundamental interactions in nature: electromagnetic, weak and strong, with gauge group $U(1)$, $SU(2)$ and $SU(3)$, respectively. Only gravity has evaded a consistent quantum gauge formulation. Thus, LGTs offer a new way to tackle hard problems in gauge theories, either by analytical or by numerical calculations.
The most prominent example is quark confinement, which has been shown to exist in a 4D $SU(3)$ LGT, but it is still unclear whether it survives in the continuum limit~\cite{Cl}.

Generally, LGTs can be classified according to their gauge group: Abelian discrete LGTs, with gauge group $\mathbb{Z}_q$, Abelian continuous LGTs, with gauge group $U(1)$, non--Abelian discrete LGTs, with the permutation group S$_n$, and non--Abelian continuous LGTs, with gauge group $SU(n)$. These theories have applications in a variety of fields of physics. We have already mentioned applications of continuous LGTs in the Standard Model, when one takes the limit of continuous spacetime. Non--Abelian discrete LGTs, with the permutation group S$_3$, are used to describe magnetic monopoles and topological defects. 
In this work we will focus on \emph{Abelian discrete} LGTs. 

Despite being the simplest instances of LGTs, Abelian discrete LGTs already have numerous applications in physics. 
Already the $\mathbb{Z}_2$ LGT on a 3--dimensional square lattice (3D $\mathbb{Z}_2$ LGT) exhibits a nontrivial phase diagram with a confined and a deconfined phase (which is known from its duality relation to the 3D Ising model), thereby mimicking the confinement properties that appear in strong interactions, described by $SU(3)$ LGTs. 
On the other hand, $\mathbb{Z}_q$ LGTs can also be used as toy models to approximate $U(1)$ LGTs. To do so, one only needs to let the number of levels of every particle $q$ tend to infinity, as we will do in this work. 
A third application was introduced by 't Hooft, who observed that $\mathbb{Z}_q$ is the center of $SU(n)$~\cite{tH78}. 
Thus, gaining insight into $\mathbb{Z}_q$ LGTs may shed light on $SU(n)$ LGTs, and, as mentioned above, the latter are used to describe weak and strong interactions.
Still another application is the fact that the hamiltonian formulation of $\mathbb{Z}_q$ LGTs in $d$ dimensions can be mapped to a quantum theory in $d-1$ dimensions \cite{Ko75}.
Furthermore, Abelian discrete LGTs are studied in quantum error correction of topological quantum memories, since a faulty syndrome measurement of these memories can be mapped to a $\mathbb{Z}_q$ LGT with randomness~\cite{De02}.
Finally, $\mathbb{Z}_q$ LGTs can also be used to study spin glasses~\cite{Ni01}.

\emph{Formal definition of an Abelian discrete LGT}.
We consider a standard definition of an Abelian discrete LGT using a Wilson Hamilton function in terms of face interactions, with gauge group $\mathbb{Z}_q$. That is, classical spins sit at the edges $e\in E$ of a $d-$dimensional lattice, and they have $q$ levels, $s_e=0,1,\ldots,q-1$. They interact via the faces $f\in F$ of the lattice; thus, a face of $k$ sides reflects a $k$--body interaction. More precisely, the Hamilton function of the system reads
\begin{equation}
H(\mathbf{s}) = - \sum_{f\in F} J_f
\: \mathrm{Re}\left[ \prod_{e\in\partial f} e^{i\frac{2\pi}{q}s_e}\right],
\label{eq:H}
\end{equation}
where $J_f$ is the interaction strength at face $f$, $\mathrm{Re}$ stands for the real part of the expression, and $e\in \partial f$ refers to the spins at the boundary of face $f$. Here $\mathbf{s}:=(s_1,\ldots,s_{|E|})$ stands for the spin configuration of the system. 
For $q=2$ (and any $k$ and $d$) we will refer to these interactions as ``Ising--type interactions'', and to the model as $d$--dimensional $\mathbb{Z}_2$ LGT (dD $\mathbb{Z}_2$ LGT) (by default being defined on a square lattice, i.e.~$k=4$). Notice that in this case each face interaction takes the form $J_{1\ldots k}(-1)^{s_1+\ldots+s_k}$, that is, it only depends on the parity of the $k$ adjacent spins.
Note also that the Hamilton function~(\ref{eq:H}) corresponds to a ``pure'' lattice gauge theory (the quotations are due to this concept being usually defined for $U(1)$ LGTs or $SU(n)$ LGTs), since we only have ``gauge fields'' at the edges and there are no particles (``matter fields'') at the vertices. 

The Hamilton function~(\ref{eq:H}) is invariant under the gauge transformation
\be
g_v=\prod_{e:e\textrm{adj } v} X_e 
\ee
where $e\textrm{adj } v$ are the edges adjacent to a vertex $v$, and $X_e$ is defined as $X_e: s_e \mapsto (s_e+1)_{\textrm{mod } q}$ (i.e.~it is the classical analogue of the generalized Pauli operator $X$). The gauge group is generated by these transformations applied on any vertex $v$, i.e.~$\mathbb{Z}_q:= \langle g_v, \forall v\in V \rangle $.
LGTs are defined on an oriented lattice, that is, a lattice in which each edge has a \emph{tail} (one of its end vertices) and a \emph{head} (the other end vertex). Then one applies the gauge operation to an edge adjacent to $v$ if $v$ is the tail of that edge, and the inverse gauge operation if $v$ is the head of edge, where this convention is arbitrary. In this manner, every closed face (i.e. where all edges are oriented either clockwise or counterclockwise) will be gauge invariant, since a gauge operation in any of its vertices will cancel.
 In fact, one can assign a different orientation to the same edge for each face where it participates (see, e.g., \cite{Ko79} p.~694), since it is only relevant that every face is closed.
Note that only in $\mathbb{Z}_2$ LGTs it is irrelevant to have an oriented lattice, since every element in the gauge group corresponds to its own inverse. 

In our proof of the main result, we will make use of the \emph{gauge fixing} of edges. This consists of fixing spin values to zero at the expense of reducing the gauge symmetry of the model. The resulting model is physically equivalent to the original one as long as one fixes spins at edges forming at most a maximal tree (i.e.~not forming a closed loop~\cite{Cr77}). In other words, all these models would be described by the same ``effective'' Hamilton function (i.e.~acting on the actual degrees of freedom, which are those described by the Hamilton function constrained by its symmetries), and in this sense they can be regarded as belonging to the same equivalence class. However, if edges fixed by the gauge form a (closed) loop, the physics described by the model change. 
Intuitively, this can be understood in several ways. 
First, if we fix all spins around the face, then the interaction in this face is fixed ($J_f (-1)^{0}$), so this is equivalent  to deleting that face. Hence, this would amount to creating a ``hole'' in that face or changing the topology of the lattice. 
Another way to see it, is that in lattice gauge theories order parameters are products of gauge fields around a closed loop. Hence, one cannot fix all these gauge without affecting the order parameter. 

The partition function of an Abelian discrete LGT is defined as
\begin{equation}
Z_{\textrm{LGT}}:=\sum_{\mathbf{s}}e^{-\beta H(\mathbf{s})},
\label{eq:Z}
\end{equation}
where $\beta = 1/(k_B T)$ is the inverse temperature. Although the partition function has no physical interpretation as such, it is important because one can compute all other thermodynamical quantities as a function of it~\cite{Pa}. For example, the mean energy is obtained by taking the derivative with respect to $\beta$ of the logarithm of $Z$ ---we will give examples of these calculations in Sec.~\ref{sec:computing-observables-complete}. 
 Thus, complete knowledge of the partition function as a function of the different variables (such as temperature, volume, number of particles, etc) amounts to complete knowledge of the system (with a fixed number of particles, since it is the canonical partition function) in thermal equilibrium.
Note that the computation of the partition function is usually regarded as a hard problem, since it involves the sum of an exponential number of terms.

To have an intuitive picture of a gauge symmetry, one can think of every lattice site as the origin of a coordinate system~\cite{Ko75}. Then, the physics of, say, a rigid rotator should be independent of whether it is described from the space coordinate system, namely, the coordinate system at site $i$, or from the body coordinate system, namely, the coordinate system at site $j$, where $(i,j)$ are linked by an edge. In this analogy, the gauge field that is attached, by definition, to the edge $(i,j)$ corresponds to a ``rotation matrix'' in going from one coordinate system to the other. Then, rotating the local coordinate system at $i$ corresponds to applying a gauge transformation at $i$, which affects all gauge fields at edges adjacent to vertex $i$. Similarly, fixing the orientation of the coordinate system at $i$ with respect to coordinate system at $j$ corresponds to fixing the gauge field at edge $(i,j)$. 

\emph{Comparison with an SSM.} 
In order to illustrate what a global symmetry is, we consider the Hamilton function of an Ising model:
\be
H_{\mathrm{Ising}}=-J\sum_{<i,j>}s_is_j, 
\label{eq:H-Ising}
\ee
where the spins take values $s_i=\pm 1$ [note that this is different from Eq.~(\ref{eq:H})], and the sum is over nearest neighbors. Here it is clear that a global flip of all spins leaves the Hamilton function invariant, whereas any local spin flip would amount to some energy change. 

\section{Completeness of the 4D $\mathbb{Z}_2$ LGT}
\label{sec:completeness-4D-Z2-LGT}

In this section we prove the main result of this paper, namely the completeness of the 4D $\mathbb{Z}_2$ LGT. This means that the partition function of any Abelian discrete classical spin model (including $\mathbb{Z}_q$ LGTs as well as discrete SSMs) equals the partition function of an \emph{enlarged, inhomogeneous} 4D $\mathbb{Z}_2$ LGT. The couplings of the 4D $\mathbb{Z}_2$ LGT are precisely the parameters that have to be tuned so that its partition function equals one specific target model and not another.

As mentioned in the introduction, similar results were obtained in \cite{Va08,De08}, but they either relied on having imaginary couplings~\cite{Va08}, or where much more restricted~\cite{De08}. The result presented here is both general and relies on real couplings. Although it was essentially given in \cite{De09}, here we will present it in greater detail and rigor, as specified in the introduction.

In order to prove the main result, we will proceed as follows:
\begin{enumerate}
\item[(i)] In Sec.~\ref{ssec:quantum-formulation} will present a quantum formulation of the partition function of all $\mathbb{Z}_q$ LGTs;
\item[(ii)]
In Sec.~\ref{ssec:merge-deletion-rules} we will present the merge and the deletion rule, which are the two tools that will enable us to transform one interaction pattern into another;
\item[(iii)]
In Sec.~\ref{ssec:method} we will prove that the Hamilton function of an interaction pattern made of all $k$--cliques of $n$ 2--level particles, for $k=0,1,\ldots,n$, (the \emph{superclique}), with appropriate coupling strengths, equals a general Hamilton function on these $n$ particles;
\item[(iv)] 
In Sec.~\ref{ssec:explicit-construction} we will use the merge and deletion rule, as well as the gauge fixing of edges, to construct a superclique out of a 4D $\mathbb{Z}_2$ LGT;
\item[(v)] 
In Sec.~\ref{ssec:main-result} we state the main result of this paper, namely the completeness of the 4D $\mathbb{Z}_2$ LGT. This follows from the construction of the superclique (Sec.~\ref{ssec:explicit-construction}), and from the equivalence of the superclique to any Hamilton function (Sec.~\ref{ssec:method}). 
\end{enumerate}

Then, we will discuss the following aspects of our result:
\ben
\item[(vi)]
In Sec.~\ref{ssec:efficiency} we investigate how the system size of the 4D $\mathbb{Z}_2$ LGT scales with the size of the target model, and show that a polynomial overhead is required in all relevant cases;
\item[(vii)]
In Sec.~\ref{ssec:approximate-completeness} we show that the completeness of the  4D $\mathbb{Z}_2$ LGT also holds approximately for continuous models, including all $U(1)$ LGTs as well as (Abelian) continuous SSMs;
\item[(viii)]
Finally, in Sec.~\ref{ssec:space-of-all-theories}, we discuss our main result in terms of the space of all theories.
\een

\subsection{Quantum formulation}
\label{ssec:quantum-formulation}

Here we present a quantum formulation of the partition function (\ref{eq:Z}). Similar formulations have been presented in \cite{Va08,Va07}, and have proven to be useful, inasmuch as they have used quantum mechanical tools to gain insight into SSMs. The basic idea is that manipulations of the quantum states that appear in the quantum formulation of a model allow one to transform them into the quantum formulation of another model. Via this detour into quantum mechanics, one actually maps the partition function of the former model into the partition function of the latter.

\begin{figure}[htb]
\centering
\includegraphics[width=1\columnwidth]{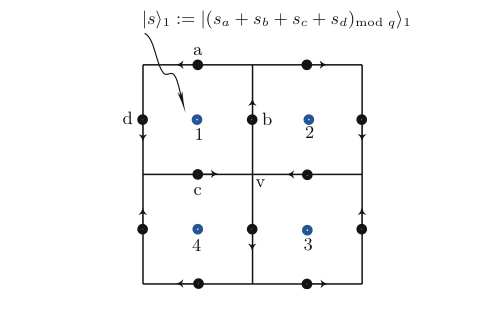}
\caption{The state $|\psi_{\mathrm{LGT}}\ra$ places one quantum particle (blue dots, labeled with numbers) at each face, thereby characterizing the interaction of classical spins (black dots, labeled with letters) around that face. The quantum spin at face $f_1$, $s_{f_1}$ is obtained by summing modulo $q$ the values of the spins at the boundary of the face. }
\label{fig:adLGT-graphpicture}
\end{figure}

We define an (unnormalized) quantum state of $|F|$ $q-$level quantum particles: 
\be
|\psi_\mathrm{LGT}\ra = \sum_{\mathbf{s}} \bigotimes_{f\in F} |(\sum_{e\in\partial f} s_e)_{\textrm{mod } q}\ra_f
\label{eq:psiLGT}
\ee
where the sum is taken over all spins at the edges $e$ which are at the boundary of $f$, $\partial f$. That is, one quantum system is placed at every face of the lattice in order to characterize the interaction at that face (see Fig.~\ref{fig:adLGT-graphpicture}).
We also define an (unnormalized) complete product state 
\be
|\alpha\ra = \bigotimes_{f\in F} |\alpha\ra_f,
\label{eq:alpha} 
\ee
with 
\be
|\alpha\ra_f = \sum_{s_1,\ldots,s_k} e^{\beta J_f
  \cos\left[\frac{2\pi}{q}(s_1+\ldots+s_k)\right]} 
  |(\sum_{e\in\partial f} s_e)_{\textrm{mod } q}\ra_f .
\label{eq:alpha_f}
\ee
That is, $|\alpha \ra_f$ contains the factor with which face $f$ contributes to the partition function; we will sometimes write this state as $|\alpha (J_f)\ra_f$ to emphasize its dependence on $J_f$.
The basis states in~(\ref{eq:psiLGT}) and~(\ref{eq:alpha_f}) are the eigenstates of the quantum phase shift operator $Z|j\ra := e^{i2\pi j/q}|j\ra$, for $j=0,1,\ldots,q-1$.

Then, the quantum formulation of the partition function of the Abelian discrete LGT (\ref{eq:Z}) is obtained by
computing the scalar product between $|\psi_\mathrm{LGT}\ra$ and $| \alpha \ra$:
\begin{equation}
Z_\mathrm{LGT} = \la \alpha | \psi_\mathrm{LGT}\ra .
\label{eq:ZLGT}
\end{equation}
The proof of the previous equality is straightforward, as one simply needs to write down explicitly the quantity on the right hand side (r.h.s.) of (\ref{eq:ZLGT}) and see that it coincides with the partition function $Z$.
Notice that $|\psi\ra$ contains the information of the interaction pattern of the model (i.e.~the lattice on which it is defined), whereas $|\alpha\ra$ encodes the coupling strengths and the temperature of the model. Note also that the normalization of the states $|\psi_\mathrm{LGT}\ra$ and $| \alpha \ra$ would amount to an additional prefactor in the equality (\ref{eq:ZLGT}), which would not change the physics described by that partition function.

\emph{Stabilizer state.}
The state $|\psi_{\mathrm{LGT}}\ra$ [Eq.~(\ref{eq:psiLGT})] is a stabilizer state~\cite{Go97,Va05}, since it can be rewritten as
\begin{equation}
|\psi_{\mathrm{LGT}}\ra = \sum_{\mathbf{s}}|A \mathbf{s}\ra ,
\label{eq:stabilizer-state}
\end{equation}
where   $A$ is a $|F|\times |E|$ matrix whose entry $(f,e)$ is one if $e$ is a boundary edge to the face $f$, and it is zero otherwise. 
In a $d$--dimensional square lattice with periodic boundary conditions, each face has four boundary edges, and each edge is boundary to $2(d-1)$ faces. Thus, in this case, each 
row (face) contains four ones and the rest are zeros, and each column (edge) contains $2(d-1)$ ones. In the case of open boundary conditions, the edges at the boundary of the lattice are boundary to only $(d-1)$ faces, whereas all faces still have four boundary edges. 

In order to find the generators of the stabilizer of a state such as \eqref{eq:stabilizer-state}, one does Gaussian elimination to find a maximum set of linearly independent columns of $A$, say $k$. These are used to construct the generators with $\sigma_x$ operators of the stabilizer. Then one generates $n-k$ linearly independent vectors to all the rest (where $n$ is the number of particles of the state), which are used to construct the generators with $\sigma_z$ operators of the stabilizer group. 

Note that in the case of the state~\eqref{eq:stabilizer-state}, the rank of the matrix $A$ depends on the dimension of the square lattice $d$, i.e. this state will be different depending on the dimension of the lattice on which it is defined.
More precisely, one can see that the state $|\psi_{\mathrm{LGT}}\ra$ defined on a 2D square lattice with open boundary conditions corresponds to the product state $|+\ra^{\otimes |F|}$, where $|+\ra$ is the eigenstate of the Pauli matrix $\sigma_x$, with eigenvalue +1, $\sigma_x|+\ra = |+\ra$.  On the other hand, $|\psi_{\mathrm{LGT}}\ra$ defined on a 2D square lattice with periodic boundary conditions corresponds to the state $|GHZ\ra = |0\ra^{\otimes |F|} + |1\ra^{\otimes |F|} $. For higher dimensional lattices, e.g. 3D lattices, the state $|\psi_{\mathrm{LGT}}\ra$ is less trivial.

Below, we will use the state $|\psi_{\mathrm{LGT}}\ra$ defined on some lattice as a resource for measurement--based quantum computation \cite{Ra01} in order to prove the completeness of that LGT on that lattice. The fact that $|\psi_{\mathrm{LGT}}\ra$ defined on a 2D lattice contains either no entanglement at all (open boundary conditions) or a very small amount of it (periodic boundary conditions), and thus are useless states from the point of view of measurement--based quantum computation, is in agreement with the fact that 2D $\mathbb{Z}_2$ LGTs are trivial \cite{Ko79} and cannot be complete. 

As noticed in \cite{Va07}, the fact that $|\psi_{\mathrm{LGT}}\ra$ is a stabilizer state reveals  some symmetries in the partition function. That is, because $|\psi\ra$ is left invariant under any operator $s\in \mathcal{S}$, $s|\psi\ra =|\psi\ra $, this translates into  the following invariance in the partition function
\be
Z= \la \alpha |\psi\ra  = \la \alpha |s|\psi\ra.
\ee
This implies that there is another set of couplings, determined by $\la \alpha ' |= \la \alpha | s$ that yields the same partition function as the original set $\la \alpha |$. 

\subsection{Merge and deletion rules}
\label{ssec:merge-deletion-rules}

\begin{figure}[htb]
\centering
\includegraphics[width=1\columnwidth]{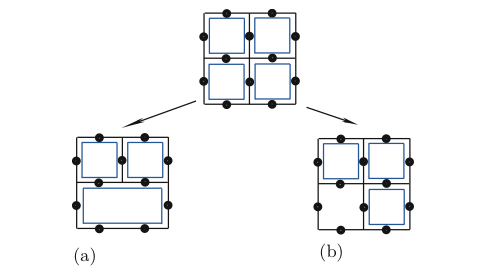}
\caption{(a) The merge and (b) deletion rules are tools that allow us to map one interaction pattern to another. Blue lines indicate faces where there are interactions}
\label{fig:manipulation}
\end{figure}

We now present two rules, the merge and deletion rule, which allow us to manipulate the partition function of a model and relate it to the partition function of another model. The intuitive picture is that the merge rule applied to a model with, say, 4 faces, transforms it to a model with 3 faces, one being larger, and containing a 6--body instead of a 4--body interaction (see Fig.~\ref{fig:manipulation}(a)). And applying the deletion rule to a face amounts to mapping it to a model where there is no such face (see Fig.~\ref{fig:manipulation}(b)). 
Although these rules can be generally defined for $\mathbb{Z}_q$ LGTs, we will henceforth focus on the case $\mathbb{Z}_2$, since this is what we require for the proof.

\begin{figure}[htb]
\centering
\includegraphics[width=1\columnwidth]{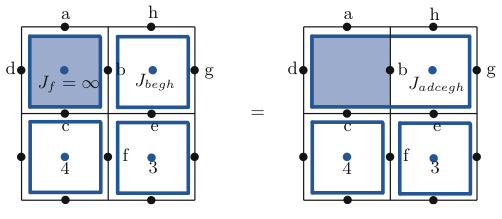}
\caption{Merge rule. Setting $J_{f}=\infty$ sets the condition $s_a+s_b+s_c+s_d=0$, and thus, one of the variables becomes dependent, say $s_b= s_a+s_c+s_d$. This is substituted in the right face, which now depends on $s_h+s_g+s_e+s_a+s_c+s_d$, i.e.~a 6--body Ising--type interaction, with the coupling strength of the enlarged face, $J_{begh}$ (now $J_{adcegh}$).}
\label{fig:mergerule}
\end{figure}

\emph{Merge rule.}
 The rule works by setting the coupling strength of a face, say $J_f$, to infinity. In order to see its effect, we consider (\ref{eq:alpha_f}), and we divide each coefficient by a factor of $e^{\beta J_f}$,
\begin{equation}
|\alpha\ra_f = \sum_{s_1,\ldots,s_k} e^{\beta J_f
 ( \cos\left[\pi(\sum_{e\in\partial f} s_e)_{\textrm{mod } 2}\right] -1 )} 
 |(\sum_{e\in\partial f} s_e)_{\textrm{mod } 2}\ra_f .
\label{eq:alpha_f-rescaled}
\end{equation}
Since this is a rescaling of the energy, this does not modify the relevant physics that one can derive from the partition function. In Eq.~(\ref{eq:alpha_f-rescaled}) it is clear that when $J_{f}\to\infty$, only the coeffcient with 
\be
\cos\left[\frac{2\pi}{q}(\sum_{e\in \partial f} s_e)_{\textrm{mod }2}\right]=1,
\label{eq:condition-mergerule}
\ee
 i.e.~$(\sum_{e\in \partial f} s_e)_{\textrm{mod }2}=0$ remains non-zero. That is, the overlap with $|\alpha_f(J_{f}=\infty)\ra $ becomes a projection onto the $|0\ra_{f}$ state, and imposes the condition $(\sum_{e\in \partial f} s_e)_{\textrm{mod }2}=0$ on the remaining terms. Due to this condition, one of the spins around $f$ is not free anymore, but equals the sum of the other $k-1$ spins (since it is mod 2), say $s_b = s_a+s_c+s_d$ in Fig.~\ref{fig:mergerule}. This condition is substituted in another face where $s_b$ participates, e.g.~in Fig.~\ref{fig:mergerule}, the face depending on $s_h+s_g+s_e+s_b$ becomes $s_h+s_g+s_e+s_a+s_c+s_d$. Thus, this effectively enlarges the face, that is, two 4--body Ising--type interactions have become one 6--body Ising--type interaction by means of the merge rule. Note that this remaining 6--body interaction has a coupling strength given by the face which has been enlarged (see Fig.~\ref{fig:mergerule}).

The concatenation of merge rules and the gauge fixing of some particles allow us to achieve $k$--body Ising interactions, for any $k$. For example, in order to generate a 5--body interaction, we would apply the same process as in Fig.~\ref{fig:mergerule}, and we would gauge fix one of the spins at the boundary, say $s_a$.

\begin{figure}[htb]
\centering
\includegraphics[width=1\columnwidth]{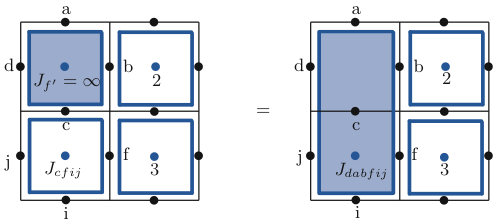}
\caption{Once we set $J_{f}=\infty$ it is arbitrary to choose in what direction this face is merged. Here, $s_c$ is chosen to be the dependent variable, and thus $f$ is merged downwards.}
\label{fig:mergerule2}
\end{figure}

Note that selecting what particle on the boundary of $f$ is dependent on the others is an arbitrary choice. That is, the face $f$ in Fig.~\ref{fig:mergerule} could have been merged with the lower or right face ($s_c$, $s_b$ dependent, respectively; see Fig.~\ref{fig:mergerule2}), or with other faces if had more than 2 dimensions. However, all choices yield equivalent partition functions. 

 We also remark that using the merge rule to transform a $k$--body interaction to a $k'$--body interaction, with $k'> k$, is only possible if $k\geq 3$, since 
 \be
 k'=2k-2,
 \label{eq:k'-k}
 \ee
(In the case of $k=2$ the spins would we sitting in the vertices and the interactions would be through the edges; however, the argument still holds true: applying the merge rule along an edge simply creates more 2--body interactions).

\begin{figure}[htb]
\centering
\includegraphics[width=1\columnwidth]{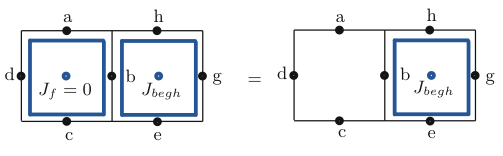}
\caption{Deletion rule. Setting $J_{f}=0$ deletes the interaction in that face.}
\label{fig:deletionrule}
\end{figure}

\emph{Deletion rule.} This rule is obtained by setting $J_{f}=0$, that is, by deleting the interaction at face $f$ (Fig.~\ref{fig:deletionrule}). Note that this corresponds to projecting the face $f$ onto the state $|+\ra=|0\ra + |1\ra$, i.e.~$|\alpha_{f} (J_f=0)\ra = |+\ra$.

\subsection{Method to obtain general $n-$body interactions}
\label{ssec:method}

Here we show that a totally general interaction between $n$ 2--level particles can be generated if all  $k$--body Ising--type interactions between these $n$ particles are available, for any subset of $k$ particles, and all $k=0,1,\ldots,n$. An interaction pattern between $n$ particles with all possible $k$--body interactions is called a $k$--clique. We coin the term \emph{superclique} for an interaction pattern between $n$ particles containing all $k$--cliques, for all $k=0,1,2,\ldots,n$~\cite{superclique}. A ``superclique of Ising--type interactions'' is a superclique such that all its interactions are Ising--type; in this work, when we refer to a superclique, we will mean this kind of superclique.
Hence, we claim that a totally general interaction between $n$ particles can be generated by preparing a superclique of Ising--type interactions among them, and by tuning its coupling strengths appropriately.

In order to prove the claim, first note that a general interaction between $n$ spins corresponds to assigning a different energy $\lambda_{\mathbf{s}}$ to each spin configuration $\mathbf{s}$. Let us indicate with a subindex on the coupling strength which particles participate in a given interaction of the superclique; e.g.~$J_{123}$ is the coupling strength of the 3--body interaction between $s_1,s_2$ and $s_3$.
Hence, we need to show that the coupling strengths in the superclique can always be tuned so that 
\bea
&&J_0(-1)^{0} + \nn\\
 &&J_1(-1)^{s_1}+\ldots+J_n(-1)^{s_n}+\nn\\
&&J_{12}(-1)^{s_1+s_2}+\ldots + J_{n-1,n}(-1)^{s_{n-1}+s_n}+\nn\\
&&J_{123}(-1)^{s_1+s_2+s_3}+\ldots + J_{n-2,n-1,n}(-1)^{s_{n-2}+s_{n-1}+s_n}+\nn\\
&&\vdots +\nn\\
&& J_{12\ldots n}(-1)^{s_1+\ldots+s_n} = \lambda_{\mathbf{s}}
\label{eq:J-lambda}
\eea
is satisfied for arbitrary $\lambda_{\mathbf{s}}$ and for all $\mathbf{s}$. 

We remark that the number of parameters is commensurate 
since there are as many $\lambda$'s as spin configurations, thus $2^n$, and as many $J$'s as interactions in the superclique, 
\be
\sum_{k=0}^n
\left(
\begin{array}{c}
n\\
k
\end{array}
\right) = 2^n\, .
\ee
Note that this includes a ``zero--body interaction'', which is a global factor $J_0$ that corresponds to a shift of all energies, and thus does not change the physics of the model. Since such a ``zero--body interaction'' cannot be prepared in the superclique, we will obtain the partition function of the final model up to this factor.  

By defining $\vec{J}$ as a column vector with all coupling strengths, 
\begin{equation}
\vec{J} =(J_0,J_1,\ldots,J_n,J_{12},\ldots, J_{n-1,n},J_{123},\ldots, J_{1\ldots n})^T,
\end{equation}
 and $\vec{\lambda}$ as a column vector with one energy $\lambda_{\mathbf{s}}$ for each column,
 \begin{equation}
 \vec{\lambda} = (\lambda_{(0,\ldots,0)}, \lambda_{(0,\ldots,1)},\ldots, \lambda_{(1,\ldots,1)})^T
 \end{equation}
 we can rewrite condition~(\ref{eq:J-lambda}) as 
 \begin{equation}
 C \vec{J} = \vec{\lambda},
 \end{equation}
  where $C$ is a square matrix with the coefficients of equation (\ref{eq:J-lambda}). That is, every row of $C$ is made of the factors $(-1)^{s_0},(-1)^{s_1},\ldots,(-1)^{s_n},\ldots, (-1)^{s_1+\ldots +s_n}$, and there is one row for each spin configuration (e.g.~the first row corresponds to $\mathbf{s} =(0,0,\ldots,0)$ and thus contains  $(-1)^{0},(-1)^{0},\ldots,(-1)^{0},\ldots, (-1)^{0}$; the second row corresponds to $\mathbf{s}=(0,0,\ldots, 0,1)$, and so on).

Thus, we need to show that $C$ can be inverted, i.e.~that one can always find $\vec{J}$ as a function of the given energies $\vec{\lambda}$, 
\begin{equation}
\vec{J} = C^{-1}\vec{\lambda}.
\label{eq:J=Clambda}
\end{equation}
In the following we will show that $C$ cannot only be inverted, but it is also an orthogonal matrix, i.e.~$C^{-1}\propto C^T$. 

Let us denote by  $\mathbf{r}_m$ ($\mathbf{r}_n$) the $m$th ($n$th) row of $C$. 
Then we want to prove that 
\be
\mathbf{r}_m\cdot \mathbf{r}_n =0\qquad \forall n,m
\label{eq:orthogonal}
\ee
where $\cdot$ denotes scalar product. Let $r_m^{i}$ be the $i$th entry of that row (which corresponds to a given interaction $i$). We define $S$ as the set of interactions $i$ for which rows $\mathbf{r}_m$ and $\mathbf{r}_n$ have the same sign,
\be
S := \{i: r_{m}^{i}  =r_{n}^{i}\}
\ee
and $D$ as the set of interactions $i$ for which the two rows have opposite sign:
\be
D := \{i: r_{m}^{i}  \neq r_{n}^{i}\}.
\ee
It is easy to see that Eq.~(\ref{eq:orthogonal}) is only satisfied if 
\begin{equation}
|S| = |D|, 
\label{eq:|S|=|D|}
\end{equation}
where $||$ denotes cardinality. In order to prove the latter statement, we will show that each element in $S$ can be paired with one element in $D$.

Now, recall that row $\mathbf{r}_m$ is obtained by substituting the value of the spin configuration, say, $\mathbf{s}_m$ into the coefficients of Eq.~(\ref{eq:J-lambda}), and similarly for row $\mathbf{r}_n$ with the spin configuration, say, $\mathbf{s}_n$. If we compare the spin configurations $\mathbf{s}_m$ and $\mathbf{s}_n$ they will differ in some positions (at least 1 and at most $n$). Let us denote by $X$ the set of spins which have a different value in $\mathbf{s}_n$ when compared with $\mathbf{s}_m$.
We pick one element of $X$ which we denote by $x$ (that is, $x$ is one spin, for example $s_3$ which, e.g., in $\mathbf{s}_n$ it takes the value 0 and in $\mathbf{s}_m$ it takes the value 1).
If a given interaction $i$ contains $x$ (e.g., the interaction $s_1+s_2+s_3$ contains $s_3$), then we pair it with an interaction $i'$ which equals $i$ but does not contain $x$ (e.g. with $s_1+s_2$). And, conversely, if $i$ does not contain $x$ (e.g.~$s_1+s_2$), then we pair it with $i'$ which equals $i$ but also contains $x$ (e.g.~$s_1+s_2+s_3$).

Let us suppose that interaction $i$ belongs to $S$. Then, interaction $i'$ has either added or removed \emph{one} element of $X$, that is, one element that has a different sign in $\mathbf{s}_n$ compared to $\mathbf{s}_m$. Thus $i'$ will belong to $D$. Similarly, if $i$ initially belonged to $D$, $i'$ will belong to $S$. Since this argument holds for all interactions $i$, we have proved that we can pair each element in $S$ with one unique element in $D$. Thus, the two sets must have equal size, viz.~Eq.~(\ref{eq:|S|=|D|}) holds.

Since the above argument holds for any pair of rows $\mathbf{r}_n,\mathbf{r}_m$, this shows that Eq.~(\ref{eq:orthogonal}) holds and that $C$ is an orthogonal matrix. This concludes the proof.

Let us remark here that in a superclique each particle participates in all possible $k$--body interactions, for any $k=0,1,\ldots,n$, that is, in
\be
\sum_{k=0}^{n-1}
\left(
\begin{array}{c}
n-1\\
k
\end{array}
\right)
=2^{n-1}
\label{eq:int-of-each-spin}
\ee
interactions~\cite{typoPRL}. This fact will be important in Sec.~\ref{ssec:explicit-construction}, where it will be the reason to require a 4D instead of a 3D lattice.

We stress that this mapping is at the level of the Hamilton function, in contrast with the other mappings used for the completeness results, which are at the level of the partition function. In plain words, we have proven that the Hamilton function of a totally general interaction of $n$ 2--level particles (e.g.~including complicated many--body interactions, etc) equals the Hamilton function of a complicated interaction pattern (a superclique), but with simple interactions (Ising--type interactions).
This mapping may have applications in the Hamiltonian formulation of LGTs~\cite{Ko75}
 and in their renormalization group analysis~\cite{Wi75}.

\subsection{Explicit construction of the superclique}
\label{ssec:explicit-construction}

In the following we show that we can construct a superclique of Ising--type interactions from a 4D $\mathbb{Z}_2$ LGT. Because of the result of the previous section, this means that, by tuning the coupling strengths of the Ising--type interactions in the superclique, this model can specialize to any other (Abelian, discrete) classical spin model. 

In order to generate the superclique starting from the 4D $\mathbb{Z}_2$ LGT we will only make use of the merge and deletion rule, and of the gauge fixing. 

We will first show the generation of $k$--body Ising--type interactions, for any $k=1,\ldots,n$ in a 3D $\mathbb{Z}_2$ LGT. Then we will argue that the fourth dimension is needed to replicate the spins, and thereby let each spin participate in all the interactions required in the superclique.

\begin{figure}[htb]\centering
\includegraphics[width=1\columnwidth]{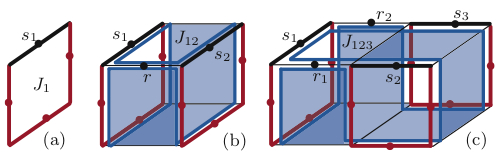}
\caption{``Logical'' spins (i.e.~the spins that will participate in the final superclique) are marked in bold black.  Blue shaded faces indicate merged faces as in Fig.~\ref{fig:mergerule}, and spins fixed by the gauge are marked in red. Figures (a), (b) and (c) show a single--body, a 2--body and a 3--body Ising--type interaction with coupling strengths $J_1$, $J_{12}$ and $J_{123}$ respectively. Spin $r$ does not participate in the interaction because the blue face depends on it twice, i.e.~$r+r=0$, and the same holds for $r_1$ and $r_2$.
}
\label{fig:1-2-3-body-interaction}
\end{figure}

First, a ``single--body'' Ising--type interaction of $s_1$ (analogous to a magnetic field), i.e.~$J_1(-1)^{s_1}$, is obtained by letting $s_1$ interact with all other spins around a face fixed by the gauge (see Fig.~\ref{fig:1-2-3-body-interaction}(a)). 

\begin{figure}[htb]\centering
\includegraphics[width=1\columnwidth]{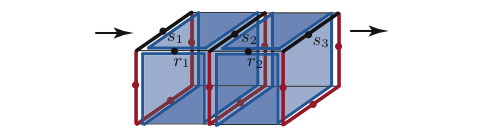}
\caption{Propagation of the spin $s_1$ to the spin $s_3$ by concatenating 2--body interactions with $J_{12}=\infty$. Particle $s_1$ has two ends (i.e.~faces that can participate in a $k$--body interaction): itself, and the right face where $s_3$ is.}
\label{fig:concatenation}
\end{figure}

\begin{figure}[htb]\centering
\includegraphics[width=1\columnwidth]{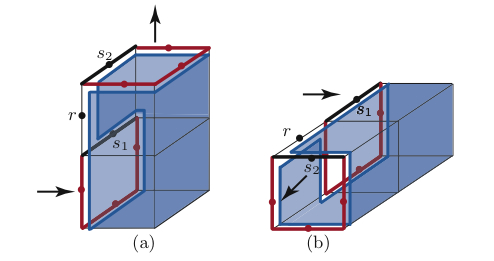}
\caption{(a) Turn in the path from a left--to--right propagation (``incoming'' spin $s_1$) to a down--to--up propagation (``outgoing'' spin $s_2$). (b) Similar turn, but here  $s_2$ propagates from back to front. The dependence of each blue face on $r$ cancels because it depends twice on it. }
\label{fig:propagation-turns}
\end{figure}

A 2--body Ising--type interaction is obtained by concatenating the merge rule on the front, lower and back face of a cube and creating the face with blue boundaries of Fig.~\ref{fig:1-2-3-body-interaction}(b). This face depends on all of the spins at its boundary (i.e. all spins which are attached to the thick, blue line in Fig.~\ref{fig:1-2-3-body-interaction}(b), e.g. the upper, right, and left spins on the front face, the lateral spins on the lower face, etc). However, six of these spins are fixed by the gauge (red spins in Fig.~\ref{fig:1-2-3-body-interaction}(b)), that is, their value is fixed to zero. Hence, the big blue face only depends on the spins which are not fixed, that is on $s_1+s_2+r+r=s_1+s_2$ (since the sum is  performed mod 2). This corresponds to the 2--body Ising--type interaction between $s_1$ and $s_2$ with coupling strength determined by the only face which has not been merged (the upper face), viz.~$J_{12}(-1)^{s_1+s_2}$. 
Furthermore, notice that by setting $J_{12}=\infty$ as well, one enforces $s_1+s_2=0$, i.e.~$s_1=s_2$. This can be seen as a ``propagation'' of the value of $s_1$ into $s_2$. A concatenated application of this 2--body interaction results in an effective propagation of a spin through a certain path in the lattice (see Fig.~\ref{fig:concatenation}). 
The direction of this propagation can be changed by merging all ``covering'' faces between the incoming and the outgoing spin, as indicated in Fig.~\ref{fig:propagation-turns}. 
This will be important in the construction of the superclique, where one needs to propagate logical particles in the 4D lattice to bring to the place where the interaction occurs.

In order to generate a 3--body Ising--type interaction, one propagates three ``logical'' spins in the lattice in order to bring as close to each other as possible, with the condition that their ``red u--shapes'' are not adjacent. Then ones merges all but one of the ``cover'' faces into a large blue face as indicated in Fig.~\ref{fig:1-2-3-body-interaction}(c). This blue face now contains the interaction $J_{123}(-1)^{s_1+s_2+s_3}$, that is, a 3--body Ising--type interaction between $s_1,s_2$ and $s_3$ as required. The interaction strength $J_{123}$ is determined by the only face that has not been merged, and it is this parameter that will be tuned so that the superclique specializes to a general $n$--body interaction. Note that here the dependence on $r_1$ and $r_2$ also cancels, since the large blue face depends on each of them twice (and the sum is mod 2, thus $r_1+r_1=0$, and similarly for $r_2$).

\begin{figure}[htb]\centering
\includegraphics[width=1\columnwidth]{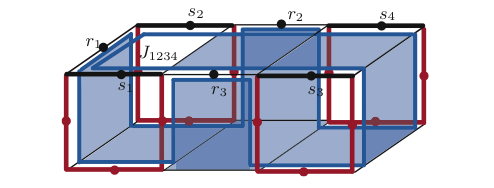}
\caption{4--body Ising--type interaction between spins $s_1,s_2,s_3$ and $s_4$ with coupling strength $J_{1234}$. See the caption of Fig.~\ref{fig:1-2-3-body-interaction} for an explanation of the symbols.}
\label{fig:4-body-interaction}
\end{figure}

\begin{figure}[htb]\centering
\includegraphics[width=1\columnwidth]{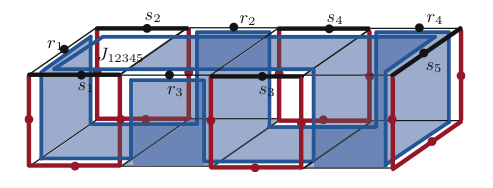}
\caption{5--body Ising--type interaction $J_{12345}(-1)^{s_1+s_2+s_3+s_4+s_5}$. See the caption of Fig.~\ref{fig:1-2-3-body-interaction} for an explanation of the symbols.}
\label{fig:5-body-interaction}
\end{figure}

4--body interactions can be generated similarly. In this case, the four logical particles are distributed close to each other as shown in Fig.~\ref{fig:4-body-interaction}, and all cover faces are merged to give rise to the Ising--type interaction $J_{1234} (-1)^{s_1+s_2+s_3+s_4}$. The procedure for the 5--body interaction analogous; in this case, one adds one more logical spin and merges the overall cover face (see \ref{fig:5-body-interaction}).

The generalization to $k$--body Ising--type interactions, for arbitrary $k$, is straightforward. First, one propagates each of the ``logical spins'' $s_1,\ldots,s_k$ in the lattice until they are as close to each other as possible, forming a ``rectangular'' shape without their red u--shapes touching each other (as is the case for the 2--, 3--, 4-- and 5--body interaction shown above). One can imagine this as adding more logical particles on the right of the 5--body interaction of Fig.~\ref{fig:5-body-interaction}, thereby enlarging the ``rectangle'' in length, until one has $k$ logical particles, analogously to how particles have been ``added'' in the generation of the 5--body interaction when compared with a 3-- or 4--body interaction. Then one merges all cover faces except for one. This renders the $k$--body Ising--type interaction $J_{1\ldots k}(-1)^{s_1+\ldots+s_k}$, where the coupling $J_{1\ldots k}$ is determined by the only cover face that has not been merged. The dependence on all auxiliary spins $r_1,r_2, \ldots$ will cancel because, by construction, the boundary of the merged face depends on them twice. The coupling strengths $J_{1\ldots k} $ will be tuned so that the Hamilton function of the superclique equals the Hamilton function of the specific final model (which we will refer to as the ``target'' model), according to Eq.~(\ref{eq:J=Clambda}).

\begin{figure}[htb]\centering
\includegraphics[width=1\columnwidth]{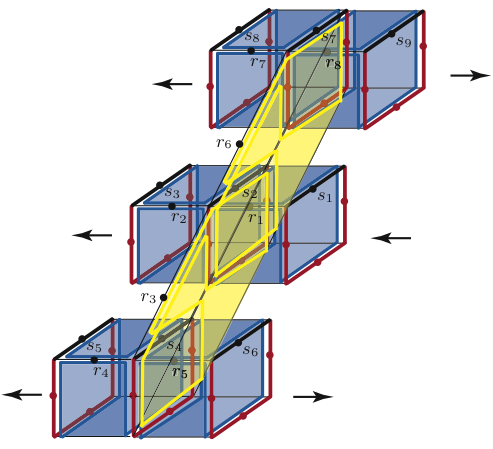}
\caption{Replication of spins in a 4D lattice: $s_1$ is replicated into $s_3,s_5,s_6,s_8,s_9$. Yellow faces have the same meaning as blue faces, that is, $s_2$ propagates into $s_3$ by the same means as it propagates into $s_7$. Note that no loops of red spins are formed. }
\label{fig:4D-replication}
\end{figure}

Thus, we have shown how to obtain $k-$body Ising--type interactions, for any $k=1,\ldots, n$.  Now we must let each spin participate in $2^{n-1}$ interactions, as pointed out in (\ref{eq:int-of-each-spin}).
However, we have seen that a spin propagates as in Fig.~\ref{fig:concatenation}, and this propagation ends in a certain face (called an ``end'') that participates in a $k$--body interaction. There it is clear that spin $s_1$ can only participate in two interactions, corresponding to the left and right ends. More generally, the number of ends that an object (or encoded particle) of dimension  $d_e$ in a lattice of dimension $d$ has are $2(d-d_e)$. Here the logical spin is never propagated alone, but always ``carries'' the other three spins of an adjacent face fixed (i.e.~the shape of Fig.~\ref{fig:1-2-3-body-interaction}(a) is propagated), hence we essentially have $d_e=2$. So, for $d=3$ the particle is blocked to have only 2 ends. We need to resort to a 4D lattice to obtain $2(d-d_e)>2$ ends (see Fig.~\ref{fig:4D-replication} for a replication in four dimensions of one spin into five other ends). Then, this replication procedure can be multiply applied until the particle has $2^{n-1}$ ends, that is, one end for each interaction. Note that in this replication procedure no loops of spins fixed by the gauge are formed.

We remark that all faces which are not mentioned in this construction have to be deleted using the deletion rule. We also mention that we have tried several other procedures in order to obtain this result in 3D, but none of them could avoid the formation of loops of edges fixed by the gauge.

The specific layout of interactions in the superclique is the following. The logical particles are distributed along the $x$ direction with one ``idle'' space among them (see Fig.~\ref{fig:superclique}). Then each of them is propagated in the $y$ direction. The idea is to use this 3D space (i.e.~the space with $w=0$) to propagate the particles, and to use the 3D space defined by $w=1$ to create the interactions required for the superclique.

\begin{figure}[htb]\centering
\includegraphics[width=1\columnwidth]{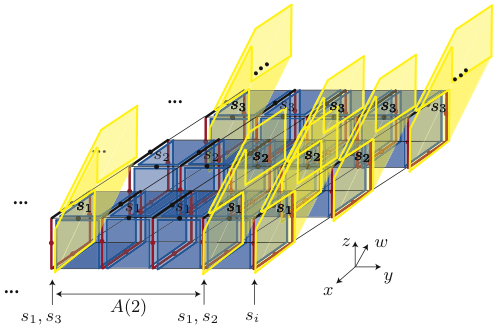}
\caption{3D space with $w=0$. Logical particles are distributed along the $x$ direction and they are propagated along the $y$ direction.}
\label{fig:superclique-1-2-3-12-13}
\end{figure}

\begin{figure}[htb]\centering
\includegraphics[width=1\columnwidth]{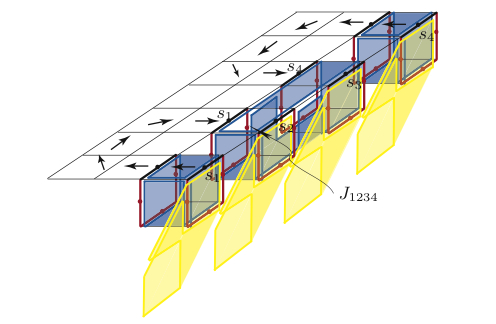}
\caption{A 4--body interaction in the superclique. The four particles $s_1$, $s_2$, $s_3$ and $s_4$ are propagated into the space with $w=1$, shown here. Here they interact in a 4--body Ising--type interaction as the one presented in Fig.~\ref{fig:4-body-interaction} with coupling strength $J_{ijkl}$. Black arrows indicate propagation of the spin (as in Figs.~\ref{fig:concatenation}, \ref{fig:propagation-turns}; the corresponding merged faces are not depicted to avoid overloading). }
\label{fig:4-clique-4-interaction}
\end{figure}

\begin{figure}[htb] \centering
\includegraphics[width=1\columnwidth]{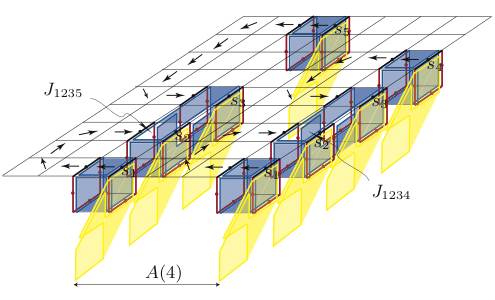}
\caption{View of part of the $w=1$ space. First, particles $s_1,s_2,s_3,s_4$ are propagated into this space, and they are brought close to each other (propagations indicated in black arrows) in order to interact in a 4--body interaction with interaction strength $J_{1234}$. After $A(4)$ idle particles in the $y$ direction, particles $s_1,s_2,s_3,s_5$ are propagated into this space and, again, they are brought close to each other to interact in the 4--body interaction with strength $J_{1235}$.}
\label{fig:4-clique-4-interaction-w1}
\end{figure}

\begin{figure}[htb]\centering
\includegraphics[width=1\columnwidth]{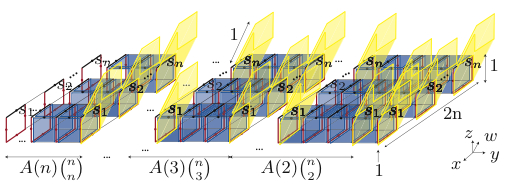}
\caption{3D space with $w=0$. Logical particles are distributed along the $x$ direction and they are propagated along the $y$ direction. A detailed part of this space is shown in Fig.~\ref{fig:superclique-1-2-3-12-13}.}
\label{fig:superclique}
\end{figure}

For example, in Fig.~\ref{fig:superclique-1-2-3-12-13} we see 3 logical particles along the $x$ direction which are propagated in the $y$ direction. Then, at some sites, they are also propagated to the space defined by $w=1$. In particular, in the first site, they are all propagated to $w=1$, where the 1--body interaction of each of the particles will take place. In the following site, $s_1$ and $s_2$ are propagated to $w=1$, in order to generate the 2--body interaction among them. After some idle propagation in the $y$ direction (precisely, $A(2)$ idle sites, as will be explained below), $s_1$ and $s_3$ are propagated to the space $w=1$, where they will generate a 2--body interaction among them. This goes on for all 2--body interactions, then for all 3--body interactions, 4--body, and so on, up to the $n$--body interaction. For example, the 4--body interaction between $s_1,s_2,s_3,s_4$ and then between $s_1,s_2,s_3,s_5$ are shown in Figs.~\ref{fig:4-clique-4-interaction}, \ref{fig:4-clique-4-interaction-w1}.


Note that the idle space one has to leave in the $y$ direction among each propagation to the $w=1$ space depends on $k$.
 This is because, when the $k$ particles are propagated to the $w=1$ space, they are distributed in a line. There, one has to rearrange them in the rectangular form explained in Figs.~\ref{fig:4-body-interaction}, \ref{fig:5-body-interaction}. 
It follows from the construction that this rearrangement requires to leave space
\be
A(k) = 2 \lceil k/4\rceil +2\: \: \sim k
\ee
between interactions. 
An overall layout of the propagation of particles in the $w=0$ space is shown in 
Fig.~\ref{fig:superclique}.

Finally, note that, as indicated in Fig.~\ref{fig:superclique}, one requires a 4D lattice of size 
\be
(x,y,z,w) = (2n, \sum_{k=0}^n A(k) 
\left(
\begin{array}{c}
n\\
k
\end{array}
\right)
,1,1) \sim 2^n 
\ee
to generate a superclique of $n$ particles. There is an exponential overhead in the system size since one has to generate an exponential number of interactions. We remark that efficient constructions can be found for specific target models, e.g.~in Sec.~\ref{ssec:magnetization} we show that the construction of the 2D Ising model only requires a linear overhead. We also point out that the construction of the 4--clique (i.e.~the part of the superclique with 4--body interactions) is the essential ingredient of the mean--field theory that we will construct in Sec.~\ref{ssec:mean-field-theory}.

\subsection{Main result}
\label{ssec:main-result}

Now we can finally gather the results we have proven in the last sections in order to state our main result. In Sec.~\ref{ssec:explicit-construction} we have shown that by setting some coupling strengths to infinity or zero (merge or deletion rule), the partition function of a 4D $\mathbb{Z}_2$ LGT can become the partition function of a superclique. Then, in Sec.~\ref{ssec:method}, we have shown that if one tunes the coupling strengths of a superclique appropriately, its Hamilton function specializes to a totally general Hamilton function between $n$ 2--level particles, and thus the corresponding partition functions are also equal. Therefore, we have shown that the partition function of an enlarged 4D $\mathbb{Z}_2$ LGT with appropriate inhomogeneous coupling strengths can specialize to the partition function of any Hamilton function of $n$ 2--level particles.
More specifically, we have seen that, for any classical spin system, there is a subsystem of the complete model (the superclique) that behaves like it (when the appropriate coupling strengths are set on it).

Let us elaborate on the class of models that are embraced by this result. First of all, the completeness result holds for models with an \emph{arbitrary interaction pattern} between these $n$ 2--level particles, which includes
\begin{itemize}
\item
Models in regular lattices in arbitrary dimension; e.g. an 8D $\mathbb{Z}_2$ LGT can be mapped to an enlarged 4D $\mathbb{Z}_2$ LGT with appropriate inhomogeneous couplings; 

\item
Models on arbitrary graphs; e.g.~a $\mathbb{Z}_2$ LGT defined on a complicated, irregular graph (note that usually Abelian discrete LGTs are only defined on hypercubic lattices and here a much more general class is considered);

\item
Models with different number of particles participating in the many--body interactions; e.g. models with 6--body interactions can be mapped to the 4D $\mathbb{Z}_2$ LGT, which has 4--body interactions. 

\item
Models with different types of many--body interactions; e.g. models containing more general 4--body interactions (the most general case being to assign a different energy to each of the 16 configurations of the 4 particles) can be mapped to the 4D $\mathbb{Z}_2$ LGT, which only contains Ising--type interactions.
\end{itemize}

In the second place, notice that the information of whether the model possesses a global or a local symmetry is also encoded in the Hamilton function. Since all Hamilton functions are included in our result, this means that the completeness result is valid for 
\begin{itemize}
\item
Models with local symmetries, i.e.~other Abelian discrete LGTs; 

\item
Models with global symmetries, i.e.~Abelian discrete SSMs; e.g.~the Ising model or the Potts model can be mapped to a model with local symmetries, the 4D $\mathbb{Z}_2$ LGT.
\end{itemize}
We will discuss how models with different types of symmetries can be mapped to each other in Sec.~\ref{ssec:magnetization}.

Furthermore, the completeness result also includes general Hamilton functions between $q-$level particles, since one just needs to encode each $q$--level particle in $m_q=\lceil \log q \rceil$ 2--level particles. Then, a totally general interaction between $n'$ $q-$level particles is generated with a superclique of $n$ 2--level particles, with $n=n'm_q$. Thus, our result also holds for 
\begin{itemize}
\item
Models whose particles have an arbitrary number of levels; e.g.~$\mathbb{Z}_q$ LGTs, which have $q$--level particles, can be mapped to the 4D $\mathbb{Z}_2$ LGT, which has 2--level particles.
\end{itemize}

In conclusion, we have shown that the 4D $\mathbb{Z}_2$ LGT is complete for all Abelian discrete classical spin models, including \emph{all} Abelian discrete LGTs and Abelian discrete SSMs. In symbols, the main result of this work can be summarized as
\be
Z_{\textrm{Abelian discrete classical}}(J) = Z_{\textrm{4D} \mathbb{Z}_2  \textrm{ LGT}} (J,J')
\label{eq:mainresult}
\ee
where $J$ is the set of couplings in the target model, and $J'$ is the set of couplings in the additional particles of the complete model.

\subsection{Efficiency results}
\label{ssec:efficiency}

We have emphasized that all completeness results require a \emph{larger}, inhomogeneous complete model when compared to the target model. Here we investigate how the number of particles in the complete model $n'$ scales with the number of particles of the target model $n$. That is, we study how the system size of the complete model increase when the system size of the target model increases.

First, we focus on the number of particles participating in interactions in the target Hamilton function. If the target Hamilton function contains at most $k-$body interactions (and $q=2$), in general one needs to generate a superclique of $k$ particles in the 4D $\mathbb{Z}_2$ LGT for each of these interactions (because the method of Sec.~\ref{ssec:method} could be applied in this case). This superclique contains $2^{k}$ Ising--type interactions, and thus the same order (up to polynomial factors) of particles in the complete model. If the target Hamilton function contains $M$ such terms, we need to generate each of these interactions, and hence require a scaling $poly(M,2^k)$.  

However, we notice that the number of particles participating in an interaction, $k$, usually does not grow with the system size. This is the case for basically all SSMs, such as the Ising, Potts or clock model, as well as vertex models (such as the 6--vertex or 8--vertex model). When this condition holds, the scaling of the complete model with $k$ is constant. We also note that for particular target models, there can be more efficient ways to create those interactions, and one does not need to create a superclique.

Next, we consider the case of $q$--state models. Each $q$--level particle requires first an encoding into $m_q=\lceil \log_2 q\rceil$ 2--level particles. Thus, a general $k$--body interaction between $q$--level particles requires to prepare a superclique of $km_q$ particles. Now the same considerations as above apply, and thus a Hamilton function of $M$ terms with at most $k$--body interactions between $q$--level particles requires an overhead 
\be
poly(M,2^{k},q)
\ee
 in the 4D $\mathbb{Z}_2$ LGT.

In summary, the target model can be prepared efficiently if $k$ scales not faster than logarithmically, and $q$ and $M$ scale polynomially with the system size.

These criteria determine whether a given continuous SSM can be approximated efficiently.
For example, an Ising model defined on a lattice has constant $q=2$ and $k=2$ for all sizes, and as the size of the lattice increases, the number of terms in its Hamilton function $M$ increases polynomially. Hence, it can be prepared efficiently from a 4D $\mathbb{Z}_2$ LGT. Even more, in Sec.~\ref{ssec:magnetization} we will present an optimized construction of the 2D Ising model which scales linearly with the system size.

Regarding Abelian discrete LGTs, they are usually defined on a hypercubic lattice, i.e.~with $k=4$ fixed, and with for some particular, fixed value of $q$. Thus, as we let the system size increase, only $M$ increases polynomially, and therefore they can be efficiently prepared from the 4D $\mathbb{Z}_2$ LGT.

\subsection{Approximate completeness for continuous models}
\label{ssec:approximate-completeness}

The completeness of the 4D $\mathbb{Z}_2$ LGT can be extended in an approximate way to continuous models. That is, the partition function of an (Abelian) continuous model can be expressed, up to a certain accuracy, as a specific instance of the partition function of the 4D $\mathbb{Z}_2$ LGT. This can be trivially seen by considering a target model with $q$--level particles, and then letting $q\to\infty$. 

As we have seen in Sec.~\ref{ssec:efficiency}, the overhead in system size of the 4D $\mathbb{Z}_2$ LGT will be polynomial as long as $q$ increases polynomially with the system size.

This extension includes continuous models with global symmetries (i.e.~continuous SSMs). For example, it includes the classical Heisenberg model
\be
H_{\mathrm{Heisenberg}}=-\sum_{<i,j>} J_{ij} \vec{s}_i \cdot \vec{s}_j
\ee
where the sum is over nearest neighbors and the spins are vectors in a unit 3--dimensional sphere, $\vec{s}_i\in \mathbb{R}^3$, $|\vec{s}_i|=1$.

The result also includes continuous models with local symmetries, i.e.~$U(1)$ LGTs, which are defined by the Hamilton function
\be
H_{U(1) \textrm{ LGT}} = -\sum_{f} J_f \cos(\sum_{e\in \partial f}\theta_e)
\ee
where the sum is over all faces $f$, and the cosine of the sum of the spins around each face is taken. Here the spin variables are angles $\theta_e\in [0,2\pi)$. 
The exploration of exact completeness results with continuous models is ongoing work.

\subsection{Space of all theories}
\label{ssec:space-of-all-theories}

In the previous sections we have established a relation that embraces models with very different features, such as different dimensions, number of levels of each particle or different types of symmetries. The purpose of this section is to picture this result in the space of all theories (to be defined below) and to compare it with other similar relations that had been obtained previously.

The \emph{space of all theories} is a space of models where the dimension of the model $d$ is depicted in one axis, the number of levels of the particles $q$ in the second axis, and the number of particles participating in an interaction $k$ in the third~\cite{Wi75}. Hence, we identify a specific model with a point in this space, which is defined by the triplet $(d,q,k)$. 
Notice, however, that a point actually corresponds to infinitely many models, since there are many features (such as the choice of parameters) that are not specified in this space, as we will point out below. 
The space of all theories is useful to visualize different universality classes, as we will elaborate on below.

\begin{figure}[htb]
\centering
\includegraphics[width=1\columnwidth]{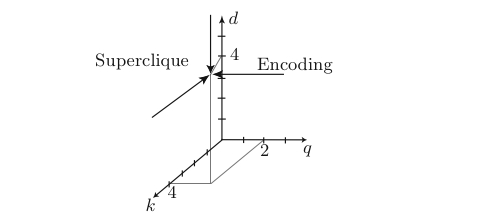}
\caption{
The main result of this paper seen in the space of all theories: models with arbitrary dimension $d$, $q$--level particles and $k$--body interactions can be mapped to the partition function of the 4D $\mathbb{Z}_2$ LGT with real parameters. The latter can specialize to models with larger $d$ or $k$  by constructing the superclique, and to models with larger $q$ by encoding $q$--level particles into 2--level particles. }
\label{fig:4DZ2LGT-complete}
\end{figure}

In the space of theories, our result can be illustrated as follows: one can map models with any $(d,q,k)$ (with inhomogeneous coupling strengths and arbitrary patterns of interactions) to a model with $(4,2,4)$ (see Fig.~\ref{fig:4DZ2LGT-complete}). Moreover, we know that this complete model has $\mathbb{Z}_2$ gauge invariance, inhomogeneous coupling strengths, and more particles than the original model (since it also has all auxiliary particles used in the construction of the superclique). 

This has been proven in different steps. We have shown that a model with fixed $(d,q,k)$ can become a model with $(d',q,k)$ for arbitrary $d'$ by preparing a $k$--clique, and it can become a model with arbitrary $(d',q,k')$ by preparing the superclique (Sec.~\ref{ssec:explicit-construction}). Then, the superclique is shown to be equivalent to a general interaction between $n$ 2--level particles (Sec.~\ref{ssec:method}). The mapping to models with general $q'$ is achieved by encoding these particles into 2--level particles. 
Note that the mapping of a model with a certain $d$ (or $q$ or $k$) to a model with larger $d$ (or $q$ or $k$) (with more particles) is trivial, since one can just leave the extra dimensions (or the extra levels or particles) empty.  

Note that, since this is only a 3--dimensional space, many properties of the models are not captured in this space, such as 
(i) the specific interaction pattern (i.e.~only the dimension of the model is specified, assuming that it is defined on a regular lattice),
(ii) inhomogeneous coupling strengths (here the coupling strength is fixed to $J$; in principle one could generalize it by adding one axis for every coupling strength $J_1, J_2,\ldots$),
(iii) whether the model has global or local symmetries, which is related to the type of interactions in the model (this could only be recognized in the universality classes of the space),
(iv) whether or not all particles of the model have $q$ levels (although it is a natural requirement to ask for the same number of levels in all particles), and (v) the number of particles of the model is not represented explicitly. 
Nevertheless, this space encodes some of the most relevant information and is useful to describe part of our main results in a graphical way. 

\begin{figure}[htb]
\centering
\includegraphics[width=1\columnwidth]{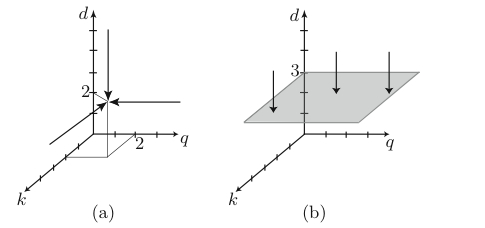}
\caption{Previous completeness results. 
(a) All models can be mapped to the 2D Ising model with inhomogeneous magnetic fields with \emph{complex} parameters \cite{Va08}.
(b) Models with large $(d,q,k)$ can be mapped to a model with $(3,q,k)$ with real parameters~\cite{De08}.
}
\label{fig:previous-complete}
\end{figure}

In order to put the present result into context, we include a figure of \cite{De08}, which represents previous completeness results in the space of all theories (Fig.~\ref{fig:previous-complete}). Fig.~\ref{fig:previous-complete}(a) represents the completeness result of \cite{Va08}, where it was proven that the partition function of all (Abelian, discrete) models can be mapped to the partition function of the 2D Ising model with inhomogeneous magnetic fields and with \emph{complex} couplings. Fig.~\ref{fig:previous-complete}(b) represents the completeness results obtained in \cite{De08}, where it was proven that the partition function of a model in $d$ dimensions, with at most $k$--body interactions and with $q$--level particles can be mapped to the partition function of a model in three dimensions, with $k$--body interactions and $q$--level particles. Despite being more restricted, the latter completeness result had the advantage of requiring only \emph{real} parameters in the complete model. For example, the result of \cite{De08} includes the fact that the partition function of a $d$--dimensional Ising model can be mapped to the partition function of a 3D Ising model with real couplings.

It is worth to point out the relation between the completeness result for SSMs and the existence of a critical dimension defined by the mean--field theory in SSMs. Namely, it is known that the critical dimension for classical spin systems in SSMs is $d=4$. For higher dimensions, a mean--field theory captures the universality properties of models. Interestingly enough, the dimension obtained with completeness results yields $d=3$, which is a theory not described by mean--field theory.
The result of Fig.~\ref{fig:4DZ2LGT-complete} is more complete in the sense that we have reduced or transformed all classical spin models into one point of the reduced space, as in Fig.~\ref{fig:previous-complete}(a), but for all theories and with real parameters.

The notion of criticality in LGTs has been less studied numerically. It is known that dD $\mathbb{Z}_q$ LGTs in a cubic lattice, with $d<4$, exhibit two phases: confined and deconfined. In the case $d\geq 4$, and $q\geq 5$, there is an additional phase, the Coulomb phase, that lies between the other two~\cite{El79,Ho79,Uk80,Fr82}.
The situation for non--Abelian LGTs is far richer but less understood.

\emph{Universality classes.} Note that this space of all theories contains models belonging to many different universality classes. That is, different families of models (each defined by a certain range of $d,q,k$), where every family behaves similarly around the critical point. Each of these models is mapped to the complete model with a specific set of parameters, and thus each of these families is mapped to the complete model with a specific regime of parameters. Thus, the phase diagram of the complete model (with one coupling strength in each axis) should contain the different universality classes, that is, it should mimic the space of all theories. 

In particular, this includes models with global symmetries. We discuss in Sec.~\ref{ssec:magnetization} how it is possible that a model with local symmetries specializes to models with global symmetries is discussed in Sec.~\ref{ssec:magnetization}.

Note that the result also includes $\mathbb{Z}_q$ LGTs with ``matter fields''. These theories are like the pure gauge theories that we have considered so far, and they additionally contain $q$--level spins at the vertices $s_i=0,1,\ldots,q-1$, for every $i\in V$. These variables transform under the gauge transformation as 
\be
g_i(s_i) = (s_i +1)_{\textrm{mod }q}.
\ee
The gauge--invariant Hamilton function of this theory is of the form
\bea
H(\mathbf{s}) &=& -\sum_{f} J_f 
\cos\left(\frac{2\pi}{q}(s_{ij} + s_{jk}+s_{kl}+s_{li})_{\textrm{mod }q}\right) \nn\\
&&-\sum_{i\to j} J_{ij}\cos\left(\frac{2\pi}{q}(- s_i+s_{ij} +s_j)_{\textrm{mod}q}\right)
\label{eq:H-matterfields} 
\eea
where $i\to j$ is an oriented edge, and $i$ ($j$) is the tail (head), and $i,j,k,l$ are the vertices at the boundary of face $f$ (note that we have changed the notation with respect to~Eq.~(\ref{eq:H}) for convenience).
We remark that this type of matter fields do not correspond to fermions since, in that case, we should introduce Grassmann variables at the site, instead of spin variables. In conclusion, we see how the partition function of an (enlarged) pure LGT (the 4D $\mathbb{Z}_2$ LGT) can also equal the partition function of an LGT with matter fields.
 Physically, this means that these matter spins can be ``reabsorbed'' into a pure LGT which is larger in size and with the proper couplings.

\section{Computing observables from the complete model}
\label{sec:computing-observables-complete}

In this section, we want to point out that our result should be used with care for the following reason. The computation of physical quantities of the target model usually requires one to take derivatives of the partition function with respect to some variable. When the partition function of the target model is expressed as a special instance of the partition function of the complete model, the complete model contains more particles, and hence more variables (e.g.~the coupling strengths of these additional particles). However, the derivatives in the target model must be taken \emph{only} with respect to the variables in the target model. This is because the couplings in the additional particles in the complete model have to be fixed, since they are used to transform the interaction pattern of the complete model to that of the target model. Therefore, they cannot be regarded as variables from the physical point of view in the target model. 

For example, the mean energy of a (target) model is obtained as \cite{Pa}
\be
U_{\mathrm{tar}} =  - \frac{\partial }{\partial \beta} \ln Z_{\mathrm{tar}}(J)
\label{eq:Utar}
\ee
We can rewrite this expression making use of our mapping, 
\be
Z_{\mathrm{tar}}(\beta J) = Z_{\mathrm{com}}(J,J') = \sum_{\mathbf{s}}e^{-\beta H_{\mathrm{tar}} (J) -\beta' H_{\mathrm{add}}(J')}
\label{eq:Ztar=Zcom}
\ee
where $J$ is the set of couplings in the target model, and $J'$ are the couplings for the additional particles of the complete model.
Then, we can use (\ref{eq:Ztar=Zcom}) to rewrite (\ref{eq:Utar}) as
\be
U_{\mathrm{tar}} = - \frac{\partial }{\partial \beta} \ln Z_{\mathrm{com}}(\beta J,\beta' J'),
\ee
where the derivative has to be taken only with respect to $\beta$, and not with respect to $\beta'$.

Other examples of thermodynamical quantities than can be obtained from the partition function are the Helmholtz free energy~\cite{Pa}
\be
A =- \frac{1}{\beta}\ln Z,
\ee
and derivatives thereof, such as the entropy $S$,
\be
S = - \left( \frac{\partial A}{\partial T}\right)_{V,N}
\ee
or the chemical potential $\mu$
\be
\mu =  \left( \frac{\partial A}{\partial N}\right)_{T,V}.
\ee
In all of these cases the derivatives should be taken \emph{only} with respect to the variables in the target model, not including the additional particles of the complete model.

In other words, there is a subsystem of the complete model which behaves as a physical system, in the sense that one can modify, say, its temperature and its physical properties vary. More precisely, this subsystem behaves like the target model. However, the rest of the complete model has a different, auxiliary status: it does not behave as a usual physical system, but its couplings and temperature must have fixed values, which are chosen such that the subsystem behaves like the target model.

\section{Applications: explicit calculations}
\label{sec:applications}

In this section we illustrate some of the uses of our main result by showing how to explicitly compute quantities of the target model as a function of the partition function of the complete model. In particular, in Sec.~\ref{ssec:Wilson-loops} we will show how to derive the order parameter of LGTs, the Wilson loop, from the partition function of the complete model, whereas in Sec.~\ref{ssec:magnetization} we will show how to compute the magnetization, which is the order parameter of some SSMs such as the Ising model, from the partition function of the complete model. In the latter section we will also discuss how a model with local symmetries (the 4D $\mathbb{Z}_2$ LGT) can specialize to a model with global symmetry (such as the Ising model).

\subsection{Wilson loops of lattice gauge theories}
\label{ssec:Wilson-loops}

A crucial quantity in an LGT is the Wilson loop, which is defined as a product of variables in the edges (gauge fields) over a closed loop $C$~\cite{Ko79}. In a $\mathbb{Z}_2$ LGT, the Wilson loop takes the form
\be
W(C)=e^{i \pi\sum_{e\in C}s_e}.
\ee
The expectation value of a Wilson loop
\be
\la W(C)\ra  = \frac{1}{Z} 
 \sum_{\mathbf{s}}e^{i \pi\sum_{e\in C}s_e}  e^{-\beta H(\mathbf{s})},
\ee
determines the phase in which the LGT is: in the confined phase $\la W(C)\ra$ decays as the area enclosed by the contour $C$, whereas $\la W(C)\ra$ decays with the perimeter of $C$ in the deconfined phase. This quantity can be derived from a partition function with sources~\cite{It89}:
\be
Z_{\mathrm{sources}}=\sum_{\mathbf{s}} e^{-\beta 
(H(\mathbf{s}) - \frac{1}{\beta}\sum_{f\in F}h_f s_f )} . 
\label{eq:Zsources}
\ee
where $s_f= e^{i\pi \sum_{e\in C}s_e}$. Specifically,
\be
\la W(C)\ra = \left[\prod_{f\in C} \frac{\partial}{\partial h_f} \ln Z_{\mathrm{sources}}\right]_{h_f\to 0}
\ee
where one takes derivatives with respect to all faces enclosed by the loop $C$.

Here $Z_{\mathrm{sources}}$ is our target partition function. That is, we would first generate a superclique from the 4D $\mathbb{Z}_2$ LGT, and then tune its coupling strengths so that its Hamilton function equals the exponent of Eq.~(\ref{eq:Zsources}). Consequently, the Wilson loop of the target model can be computed as
\be
\la W(C)\ra = \left[ \frac{\partial}{\partial h_f}\ln Z_{\mathrm{com}} (J,J')\right]_{h_f\to 0}.
\ee 

The Wilson loop is also defined for $\mathbb{Z}_q$ LGTs, and this derivation can also be carried over for that case.

\subsection{Magnetization of the 2D Ising model}
\label{ssec:magnetization}

Here we want to illustrate our mapping when we take an SSM as our target model. More precisely, we consider the 2D Ising model without fields, and we compute its magnetization from the partition function of the 4D $\mathbb{Z}_2$ LGT. The magnetization is defined as
\be
M_{\mathrm{tar}} = \left[ \frac{\partial}{\partial h}\ln Z_{\mathrm{tar}} (J,h) \right]_{h\to 0}
\label{eq:Mtar}
\ee
where 
\be
Z_{\mathrm{tar}} =  \sum_{\mathbf{s}} e^{-\beta (H_{\mathrm{tar}}(J) -(1/\beta) h\sum_i s_i) }.
\ee
Now, using Eq.~(\ref{eq:mainresult}) we can rewrite Eq.~(\ref{eq:Mtar}) as 
\be
M_{\mathrm{tar}} = \left[ \frac{\partial}{\partial h}\ln Z_{\mathrm{com}} (J,J',h) \right]_{h\to 0}
\label{eq:Mtar-Zcom}
\ee
where
\be
Z_{\mathrm{com}} =  \sum_{\mathbf{s}} e^{-\beta (H_{\mathrm{tar}}(J) + H_{\mathrm{add}}(J') -(1/\beta) h\sum_i s_i) }.
\ee
Note that here $\{s_i\}$ are the particles in the target model.
It is known that the magnetization of the 2D Ising model is its order parameter, and that it is non--zero,  $M_{\mathrm{tar}}\neq 0 $, in the so--called ordered phase~\cite{Pa}. On the other hand, Elitzur's theorem~\cite{El75} states that the magnetization of a model with local symmetries must remain zero. 

This applies, in particular, to the magnetization of our complete model, the 4D $\mathbb{Z}_2$ LGT. That is, the r.h.s.~of Eq.~(\ref{eq:Mtar-Zcom}) corresponds to the magnetization of the complete model (with couplings $J,J'$), $M_{\mathrm{com}}$, and this quantity must be always zero, $M_{\mathrm{com}}=0$.

This apparent contradiction is due to the fact that Elitzur's Theorem only holds when the couplings are finite and, in order to construct the target interaction pattern starting from the  4D $\mathbb{Z}_2$ LGT, we require to fix couplings to infinity, i.e.~$J_f'= \infty$ for some faces $f$. This effectively amounts to breaking the local symmetry on that face, since this imposes a particular value on the spins around that face [Eq.~(\ref{eq:condition-mergerule})]. This explains how the partition function of a model with local symmetries can  specialize to the partition function of a model with global symmetries: only if the former model contains ``extreme'' values of the coupling strengths ($J'_f=\infty$), can the local symmetry break spontaneously and behave as a global symmetry. 

\emph{Explicit construction of the 2D Ising model}. In Sec.~\ref{ssec:explicit-construction} we have shown the general procedure to construct an interaction pattern starting from the 4D $\mathbb{Z}_2$ LGT (construction of the superclique). This procedure can be optimized for a specific target model, reducing (drastically) the scaling of the size of the 4D $\mathbb{Z}_2$ LGT as a function of the size of the target model. In what follows we illustrate this by constructing a 2D Ising model from the 4D $\mathbb{Z}_2$ LGT in a direct way without preparing the superclique. 

The Hamilton function of a 2D Ising model is given by Eq.~(\ref{eq:H-Ising}), with the nearest neighbors defining a 2D square lattice. Here we will construct this interaction pattern directly, without preparing first a superclique. 

\begin{figure}[htb]\centering
\includegraphics[width=1\columnwidth]{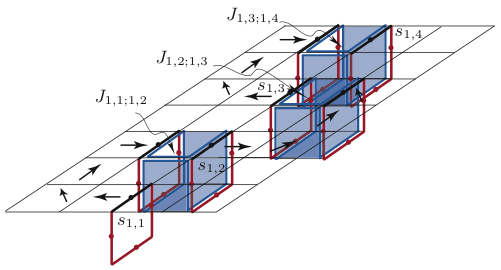}
\caption{Generation of a 1D array of Ising--type interactions: $s_{i,j}$ interacts with spin $s_{i,j+1}$ via a 2--body Ising--type interaction with coupling strength $J_{i,j;i,j+1}$. Arrows indicate propagation of the spin according to Fig.~\ref{fig:concatenation}, and the cubes marked in blue indicate a 2--body interaction according to Fig.~\ref{fig:1-2-3-body-interaction}(b).
 }
\label{fig:2DIsing}
\end{figure}

First of all, we concentrate on generating a 1--dimensional (1D) array of Ising--type interactions. In this case, we just need to generate 2--body interactions between nearest neighbors distributed on a row. This is precisely what we do on Fig.~\ref{fig:2DIsing}. There, spin $s_{1,1}$ interacts with $s_{1,2}$ in a 2--body Ising--type interaction with strength $J_{1,2}^x$, $s_{1,2}$ interacts with $s_{1,3}$ with strength $J_{2,3}^x$, and so on. The 2--body interactions are given by Fig.~\ref{fig:1-2-3-body-interaction}(b). 

\begin{figure}[htb]
\centering
\includegraphics[width=1\columnwidth]{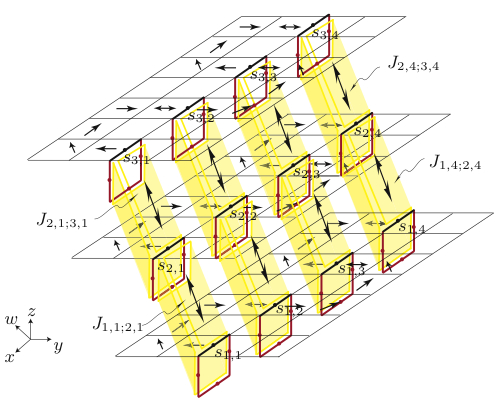}
\caption{Construction of the 2D Ising model. Each 1D array interacts with the next one via the fourth dimension, that is, $s_{i,j}$ interacts with $s_{i+1,j}$ via a yellow face, with interaction strength $J_{i,j;i+1,j}$. Every layer for different $w$ corresponds to the 1D array of interactions of Fig.~\ref{fig:2DIsing} (blue cubes are not shown to avoid overloading).
As in Fig.~\ref{fig:4D-replication}, yellow cubes have the same meaning as blue cubes, i.e.~2--body Ising--type interactions. }
\label{fig:2DIsing-4D}
\end{figure}

The next step is to create interactions among several of these 1D arrays. Since we require three dimensions to construct the 1D array of Fig.~\ref{fig:2DIsing}, we make use of the fourth dimension in order to link them as shown in Fig.~\ref{fig:2DIsing-4D}. The yellow cubes have the same meaning as the blue cubes in Fig.~\ref{fig:2DIsing}, that is, they correspond to 2--body Ising--type interactions. In this manner, spin $s_{1,1}$ interacts with $s_{2,1}$ with strength $J_{1,1;2,1}$, and so on. This completes the construction of the 2D Ising model.

As can be observed in Fig.~\ref{fig:2DIsing-4D}, the construction of a 2D Ising model of size $n \times m$ requires a 4D lattice of size $(x,y,z,w) = (2n,4,1,m)$, i.e.~the scaling is linear in the system size. 

\section{Implications of the main result}
\label{sec:implications}

In this section we will draw two implications of the main result. First, in Sec.~\ref{ssec:computational-complexity} we will conclude that computing the partition function of 4D $\mathbb{Z}_2$ LGT is $\#$P hard; that is, computationally difficult. Then, in Sec.~\ref{ssec:mean-field-theory} we will argue that our result provides a new method to compute the mean--field--theory of a $\mathbb{Z}_2$ LGT, which works for finite dimension.

\subsection{Computational complexity of 3D and 4D $\mathbb{Z}_2$ LGT}
\label{ssec:computational-complexity}

Our main result implies, in particular, that the partition function of the 2D Ising model with magnetic fields can be expressed as a specific instance of the partition function of the 4D $\mathbb{Z}_2$ LGT. The computation of the partition function of the 2D Ising model with magnetic fields is a \#P--complete problem~\cite{Ba82} --colloquially speaking, this means that it is computationally difficult~\cite{Aa}. Thus, we conclude that computing the partition function of the 4D $\mathbb{Z}_2$ LGT in the real parameter regime is \#P--hard, i.e.~at least as hard as the other problem. In other words, we have proven that one can map all models to a model which is hard to solve.
 
The construction presented above also gives insight into the complexity of the 3D $\mathbb{Z}_2$ LGT. More precisely, in Sec.~\ref{ssec:explicit-construction} we saw that a 3D $\mathbb{Z}_2$ LGT can prepare models with $k$--body Ising--type interactions, for any $k=1,\ldots,n$, as long as as every particle participates in at most two interactions (this was the limitation of the two ends of Fig.~\ref{fig:concatenation} that made us move to the 4D lattice). 
This implies, in particular, that the 3D $\mathbb{Z}_2$ LGT must be as hard as any vertex model with $q=2$ and $k-$body Ising--type interactions. 

On the other hand, using a method introduced in \cite{Va08a}, one can show that approximating the partition function of the 3D $\mathbb{Z}_2$ LGT in a certain complex parameter regime with polynomial accuracy is as hard as simulating arbitrary quantum computations, i.e.~BQP--complete~\cite{BQP}.

\subsection{Mean--Field Theory}
\label{ssec:mean-field-theory}

The mean--field theory of a model is an approximation to that model where the interaction of a variable with its neighbors is replaced by an interaction of this variable with a mean field.  
In this manner, the theory is reduced to a 1--body problem, which is useful to gain insight into a theory that is difficult to solve exactly.
Thus, there are as many ways to construct a mean--field theory of a model as ways to average over the influence of neighboring variables over a given variable.

In SSMs, mean--field theories of SSMs are generally easy to construct. For example, in the Ising model, the mean field is a mean value of the other spins, which essentially corresponds to the magnetization of the model. 

However, mean--field theories for LGTs are generally hard to construct. This is due to Elitzur's theorem~\cite{El75}, which asserts that the mean value of every variable is always zero. This problem was circumvented using a saddle--point approximation with the inverse dimension $1/d$ as an expansion parameter~\cite{Dr81,Fl82}. The restoration of the gauge symmetry in then nontrivial in this expansion~\cite{It89}.

In the following we argue that our proof of the main result yields a new method to compute the mean--field theory of a $d$ dimensional $\mathbb{Z}_2$ LGT.  This method works for fixed $d$, with $d\geq 4$, and does not break gauge invariance.

The method is based on the construction of the $4$--clique, that is, the interaction pattern in which every particle interacts with all the rest in all possible 4--body interactions. The construction of the $4$--clique for the 4D $\mathbb{Z}_2$ LGT was shown in Sec.~\ref{ssec:explicit-construction}, where we constructed a first a 1--clique, then a 2--clique, and so on for all $k$--cliques, with $k=1,\ldots,n$. Thus, the 4--clique is just one part of that construction. Since in a 4--clique every particle participates in all possible 4--body interactions, this corresponds to the coarse grained construction characteristic of a mean--field theory.

We believe that the explicit construction of the 4--clique could be useful for computer simulations of the mean--field theory of an LGT, and for this reason we presented it in detail in Sec.~\ref{ssec:explicit-construction}. We remark that the construction of a 4--clique of $n$ particles requires a 4D $\mathbb{Z}_2$ LGT of exact size
\be
(x,y,z,w)=(2n,4
\left(
\begin{array}{c}
n\\
4
\end{array}
\right)
,1,1) \sim n^5.
\ee
For example, to construct a 4--clique of 8 particles (70 interactions) a 4D lattice of size $(16,280,1,1)$ is required.

We remark that the construction of the $k$--clique requires to use the merge rule, and thus to set some coupling strengths to infinity (Sec.~\ref{ssec:explicit-construction}). As discussed in Sec.~\ref{ssec:magnetization}, Elitzur's Theorem does not apply in this regime, that is, gauge invariance is violated. Thus, in order to preserve gauge invariance, we can take a large but finite value of $J$, so that gauge invariance is still preserved. This will lead to an approximate construction of the 4--clique and thus to an approximate mean--field theory. But this does not represent a problem, since the mean--field theory is an approximate method itself. So our method allows to compute an (approximate) mean--field theory without breaking gauge invariance.

This is natural in the sense that it is at the root of the difference between SSMs and LGTs: global versus local invariance. An example may illustrate this fact. Namely, there is a complicated and indirect way to relate the 3D Ising model with the 3D $\mathbb{Z}_2$ LGT (the duality relation mentioned previously) \cite{Ko79}. Without dwelling on the details, it needs a transformation that it is non--local in the fields, even though each model is local in the fields. Therefore, it is natural that the transformation relating two types of different types of models, if exact, must be singular in the couplings. This is a way of signaling a sort of critical point separating two types of different phases or models. Our rules allow to establish a more universal and direct relation, as we have shown in the previous sections.

Note that our method requires no expansion in the dimension of the lattice $1/d$ as in \cite{Dr81,Fl82}, but it works for fixed dimension $d = 4$. It also trivially works for fixed dimension $d>4$, where one can construct the 4--clique by the same method, and simply not use the extra dimensions. Indeed, our method consists on maximally increasing the connectivity of each of the spins, which can be seen as increasing the dimensionality of the lattice (since, e.g.~in a $d$--dimensional hypercubic lattice every spin participates in $2d$ interactions). 
Note also that we can construct a $k$--clique for any $k$, and thus our method does not only apply to the usual case of a hypercubic lattice, i.e.~$k=4$, but it is more general.
Finally, notice that our construction of the $k$--clique requires a very specific gauge fixing (compare it with \cite{Fl82}, where the mean--field theory of $\mathbb{Z}_2$ LGTs is discussed with and without gauge fixing).

We mention that this construction can be generalized to $Z_q$ LGTs, since one can also define a merge rule and apply a similar construction.

\section{Extensions}
\label{sec:extensions}

In this section we present some extensions of our main result. We will argue that there are two other models (the 3D $\mathbb{Z}_2$ LGT and the 3D Ising model with 3--body interactions) which are also complete if some specific boundary conditions are imposed on them. Finally, we will compare the 3D and the 4D $\mathbb{Z}_2$ LGT and discuss whether a reduction of our result to the 3D $\mathbb{Z}_2$ LGT is possible.

\subsection{Completeness of the 3D $\mathbb{Z}_2$ LGT with fixed boundary conditions}
\label{ssec:completeness-3D-Z2-LGT-BC}

\begin{figure}[htb]\centering
\includegraphics[width=1\columnwidth]{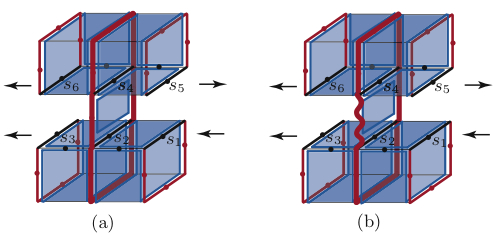}
\caption{(a) The replication of spins in a 3D $\mathbb{Z}_2$ LGT causes loops of spins fixed by the gauge (thick, red line). (b) One spin in the loop is fixed by the boundary condition (wavy, red line), thereby ``opening'' the loop of spins fixed by the gauge (straight, red lines) and allowing replication without loops in 3D. }
\label{fig:3D-replication-loops}
\end{figure}

As mentioned in Sec.~\ref{ssec:explicit-construction}, the only obstacle in proving the completeness of the 3D $\mathbb{Z}_2$ LGT was that every superclique constructed from this lattice involved loops of spins fixed by the gauge (see Fig.~\ref{fig:3D-replication-loops}), which are not allowed, i.e.~do not leave the physics of the Hamilton function invariant. An easy way to overcome this problem is to impose fixed boundary conditions in the 3D lattice. In this case, some spins are fixed to zero due to the boundary conditions, not due to the gauge fixing. Then every loop that was formed in the superclique (e.g.~the loop shown in Fig.~\ref{fig:3D-replication-loops}(a)) can be ``opened'' by replacing one spin fixed by the gauge by one fixed by the boundary condition (Fig.~\ref{fig:3D-replication-loops}(b)). 

\begin{figure}[htb]\centering
\includegraphics[width=1\columnwidth]{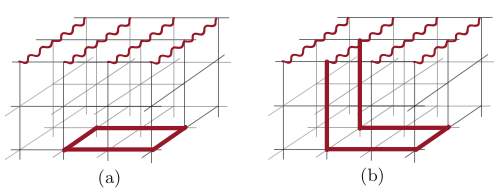}
\caption{(a) There is a loop of spins fixed by the gauge (straight, red lines) in the inner part of the lattice, and there are spins fixed by the boundary conditions (wavy, red lines) on the boundary of the lattice. (b) The loop can be ``deformed'' until one of its edges is the fixed by the boundary condition, thereby opening the loop. }
\label{fig:3D-BC2}
\end{figure}

In fact, one only needs to fix these boundary conditions at the outer boundary of the 3D lattice. This is because every loop in the inner part of the lattice (e.g.~Fig.~\ref{fig:3D-BC2}(a)) can be ``deformed'' until one of its edges becomes the edge of the outer boundary, which will open the loop (see Fig.~\ref{fig:3D-BC2}(b)).
This proves that the completeness results  of the 4D $\mathbb{Z}_2$ LGT hold as well for the 3D $\mathbb{Z}_2$ LGT with fixed boundary conditions. 

Although the bulk physics of the 3D $\mathbb{Z}_2$ LGT with boundary conditions is the same as that of the 3D $\mathbb{Z}_2$ LGT in the thermodynamic limit, in general boundary effects may give rise to new phenomena.

\subsection{Completeness of a 3D Ising model with 3-body interactions and fixed spins}
\label{ssec:completeness-3D-Ising-3-body}

We remark that the same completeness results could have been obtained with a 3D SSM with 3--body Ising--type interactions. This is because one can construct a superclique out of the this model, since 
\begin{enumerate}
\item[(i)]
this model possesses 3--body interactions, and thus the merge rule increases the number of particles participating in an interaction [see Eq.~(\ref{eq:k'-k})], and
\item[(ii)]
it is an SSM, thus the fixed spins can form loops, hence, an initial 3D lattice suffices. 
\end{enumerate}

Although this model is simpler than the 4D $\mathbb{Z}_2$ LGT (in terms of dimensions and many--body interactions) and it exhibits the same completeness property, it presents two disadvantages. The first one is that, in an SSM, all spins fixed inside the lattice must be fixed by the boundary conditions, since they do not possess any gauge symmetry. Therefore every fixed spin should be taken into account initially in the Hamilton function, which is impractical and difficult to justify physically. The second disadvantage is that it is difficult and to define a regular 3D lattice only made of 3--body interactions. Since the requirement is that the interactions are among more than two particles, instead of a lattice with only 3--body interactions, we could consider many regular triangular lattices in two dimensions, connected by parallel lines in the third dimension. 

Although an Ising model [as defined in Eq.~(\ref{eq:H-Ising}] defined on such a lattice, and with many spins fixed in the middle of the lattice, would exhibit the same completeness properties, its physical interpretation and applications are not as clear--cut as those for the 4D $\mathbb{Z}_2$ LGT.

\section{Conclusions and outlook}
\label{sec:conclusions}

We have shown that the partition function of any Abelian, discrete classical spin model can be expressed as the partition function of an enlarged 4--dimensional pure lattice gauge theory, with gauge group $\mathbb{Z}_2$, the 4D $\mathbb{Z}_2$ LGT. 
The values of the coupling strengths of the 4D $\mathbb{Z}_2$ LGT determine to what model its partition function specializes. The enlargement of the 4D $\mathbb{Z}_2$ LGT is polynomial in all relevant cases. We have also proven that the partition function of Abelian, continuous classical spin models can be expressed approximately as the partition function of the 4D $\mathbb{Z}_2$ LGT. 

We have illustrated the use of our result by computing relevant quantities of the target model as a function of the partition function of the 4D $\mathbb{Z}_2$ LGT, and by giving a specific example on how the 4D $\mathbb{Z}_2$ LGT specializes to the 2D Ising model.
We have also discussed how it is possible that a model with gauge symmetry can specialize to a model with global symmetry. 
Our result yields a new method to compute the mean--field theory for $\mathbb{Z}_2$ LGTs on hypercubic lattices of dimension $d\geq 4$. It also allows us to assert that computing the partition function of the 4D $\mathbb{Z}_2$ LGT is $\#$P--hard, that is, computationally difficult. Finally, we have shown that the same result can also be proven with two other models: the 3D $\mathbb{Z}_2$ LGT with fixed boundary conditions, and the 3D Ising model with 3--body interactions and fixed spins. 

We believe that our results render further insight into the structure of Abelian, classical spin models, as they provide an explicit mapping between very different models (including e.g. models with different types of symmetry). In particular, the fact that the phase diagram of the complete model must contain all universality classes, in principle could be used (i) as a new tool for the study of known universality classes, and (ii) as an approach to explore possibly unknown universality classes (since it embraces a larger class of models than the ones usually considered).
We also believe that our method to compute the mean field theory of $\mathbb{Z}_2$ LGTs can be useful for computer simulations.

There are many possible ways in which one could generalize and further explore our results. 
For example, it is an open question whether our result can be proven for the 3D $\mathbb{Z}_2$ LGT, instead of the 4D, since, in terms of the phase diagram, the 3D $\mathbb{Z}_2$ LGT is the simplest non--trivial model, for it exhibits a confined and a deconfined phase (this holds for cubic lattices, i.e. $k=4$).  
However, our attempts to reduce the result to 3D have failed (since loops were always formed), and we do not know if this is due to our approach or if this is not possible. One reason why this may not be possible is that in a 3D lattice there are non--trivial knots, whereas in 4D all knots are trivial~\cite{Creutz}. Thus, the loops in 3D could be seen as knots, and one may only be able to unknot them using the fourth dimension.

Another possible generalization concerns the possibility of mapping all known LGTs (including LGTs with randomness, and non--Abelian LGTs) to a certain complete LGT, from which all the rest can be derived. We have already mentioned that non--Abelian gauge groups, either discrete or continuous, are more involved and deserve a separate study. We have also mentioned that LGTs with fermionic matter should be modeled using Grassman variables and that these are beyond the scope of the mapping presented here. There is a class of random classical models, defined both in SSMs and in abelian LGTs, that has recently attracted much interest in the context of quantum error correction for topological stabilizer codes \cite{De02,Ka09,spe}.
We envisage the possibility that these models with randomness can also be fitted into a completeness construction of the type described here.

\section*{Acknowledgements}

We thank M.~Van den Nest for valuable discussions, for pointing out the connections to complexity, and for help in finding the stabilizers. 
We also thank O.~G\"uhne for very valuable help in the proof of Sec.~\ref{ssec:method}. 
This work was supported by the FWF (SFB-F40) and the European Union (QICS,  SCALA, NAMEQUAM), the Spanish MICINN grant FIS2009-10061, CAM research consortium QUITEMAD S2009-ESP-1594, European FET-7 grant PICC, UCM-BS grant GICC-910758.



\begin{thebibliography}{99}
\bibitem{typoPRL} 
In Ref.~\cite{De09} there was a typo at this point, where it was claimed that every particle must participate in $2^n/n$ interactions, instead of $2^n/2$.
\bibitem{spe}
Another argument why our result may not be reduced to a 3D $\mathbb{Z}_2$ LGT is the following. Faulty syndrome measurement in topological memories can be mapped to an LGT~\cite{De02}. 
It is known that self--correcting memories (i.e.~topological memories) can be constructed in 4D, but so far no such memory has been found in 3D~\cite{Al08,Bo09}. 
The reason for that is that the topological qubits are associated to non--trivial (i.e. homologically non--trivial) loops in the 4D surface. This protects them from $\sigma_x$ errors (bit--flip errors) as well $\sigma_z$ errors (phase--flip errors) simultaneously. In order to do this independently, one needs to locate two independent loops in a way that they do not intersect. This can be done in 4D, as one loop lives in two dimensions, and the other loop in the other two dimensions. However, this is not possible in 3D. There one can only construct a loop for the $\sigma_x$ operator and an open string for the $\sigma_z$ operator, whose boundary are two points, but not loops. Having these points is harmful and it is the reason why the information cannot be self--protected by energetic mechanisms.
\end{thebibliography}
\end{document}